\documentclass[useAMS,usenatbib]{mn2e}
\usepackage{psfrag,graphicx}
\usepackage{txfonts}
\usepackage{natbib}
\usepackage{ulem}
\usepackage{color}
\usepackage{definitions}
\usepackage{xspace}
\usepackage{subfigure}
\usepackage{hyperref}
\usepackage{textcomp}
\usepackage{graphicx,epstopdf}
\usepackage{float}

\newcommand{\reduceme}{\mbox{R\raisebox{-0.35ex}{E}D
\hspace{-0.05em}\raisebox{0.85ex}{uc}\hspace{-0.90em}
\raisebox{-.35ex}{{m}}\hspace{0.05em}E}}

\title[Abundance ratios and IMF slope in the dwarf elliptical galaxy NGC~1396 with MUSE]{Abundance ratios and IMF slopes in the dwarf elliptical galaxy NGC~1396 with MUSE \thanks{Based on observations made with ESO Telescopes at the La Silla Paranal Observatory under programme ID 094.B-0895(A)}}
\author[J.J. Mentz et al.]{J.J. Mentz$^{1,2}$\thanks{E-mail: mentz@astro.rug.nl}, F La Barbera$^{3}$, R F. Peletier$^{1}$, J.Falc\'{o}n-Barroso$^{5,6}$,  T. Lisker$^{7}$, G. van de Ven$^{8}$, 
\newauthor{S.I. Loubser$^{2}$, M. Hilker$^{9}$, R. S\'{a}nchez-Janssen$^{10}$, N. Napolitano$^{3}$},
\newauthor{M. Cantiello$^{3,4}$, M. Capaccioli$^{3}$, M. Norris$^{8}$, M. Paolillo$^{3}$, R. Smith$^{11}$, M.A. Beasley$^{5,6}$},
\newauthor{M. Lyubenova$^{1}$,R. Munoz$^{12}$, T.Puzia$^{12}$}\\\\
$^{1}$Kapteyn Astronomical Institute, University of Groningen, P. O. Box 800, 9700 AV Groningen Netherlands\\
$^{2}$Centre for Space Research, North-West University, Potchefstroom 2520, South Africa\\
$^{3}$INAF - Osservatorio Astronomico di Capodimonte, Napoli, Italy\\
$^{4}$INAF - Osservatorio Astronomico di Teramo, Italy\\
$^{5}$Instituto de Astrof\'{\i}sica de Canarias, V\'{\i}a L\'{a}ctea s/n, La Laguna, Tenerife, Spain\\
$^{6}$Departamento de Astrof\'{\i}sica, Universidad de La Laguna, E-38205, La Laguna, Tenerife, Spain\\
$^{7}$Astronomisches Rechen-Institut, Zentrum f\"{u}r Astronomie der Universit\"{a}t Heidelberg, M\"{o}nchhofstra\ss{}e 12-14, D-69120 Heidelberg,Germany\\
$^{8}$Jeremiah Horrocks Institute, University of Central Lancashire, Preston, PR1 2HE, United Kingdom\\
$^{9}$European Southern Observatory, Karl-Schwarzchild-Str. 2, D-85748 Garching, Germany\\
$^{10}$NRC Herzberg Astronomy and Astrophysics, 5071 West Saanich Road, Victoria, BC V9E 2E7, Canada\\
$^{11}$University, Graduate School of Earth System Sciences-Astronomy-Atmospheric Sciences, Seoul 120-749, Republic of Korea\\
$^{12}$Instituto de Astrof\'{\i}sica, Pontificia Universidad Cat\'{o}lica de Chile, Av. Vicu\~{n}a Mackenna 4860, 7820436 Macul, Santiago, Chile}
\begin{document}
\date{Submitted 0 August 1234; Accepted 0 October 1234}
\maketitle
\begin{abstract}
{Deep observations of the dwarf elliptical (dE) galaxy NGC 1396 (M$_V = -16.60$, Mass $\sim 4\times10^8$ M{$_\odot$}), located in the Fornax cluster, have been performed with the VLT/ MUSE spectrograph in the wavelength region from $4750-9350$ \AA{}. In this paper we present a stellar population analysis studying chemical abundances, the star formation history (SFH) and the stellar initial mass function (IMF) as a function of galacto-centric distance. Different, independent ways to analyse the stellar populations result in a luminosity-weighted age of $\sim$ 6 Gyr and a metallicity [Fe/H]$\sim$ $-0.4$, similar to other dEs of similar mass. We find unusually overabundant values of [Ca/Fe] $\sim +0.1$, and under-abundant Sodium, with [Na/Fe] values around $-0.1$, while [Mg/Fe] is overabundant at all radii, increasing from $\sim+0.1$ in the centre to $\sim +0.2$ dex. We notice a significant metallicity and age gradient within this dwarf galaxy.  

To constrain the stellar IMF of NGC 1396, we find that the IMF of NGC 1396 is consistent with either a Kroupa-like or a top-heavy distribution, while a bottom-heavy IMF is firmly ruled out.

An analysis of the abundance ratios, and a comparison with galaxies in the Local Group, shows that the chemical enrichment history of NGC 1396 is similar to the Galactic disc, with an extended star formation history. This would be the case if the galaxy originated from a LMC-sized dwarf galaxy progenitor, which would lose its gas while falling into the Fornax cluster.
}
\end{abstract}

~\hfil\break

~\hfil\break

~\hfil\break

~\hfil\break

\section{Introduction}
Knowledge of galaxy evolution relies strongly on the information obtained from studying different galaxies at various stages of their existence. Galaxy evolution depends on internal factors, as well as on external factors such as the environment and galaxy interactions \citep{Bingelli1988,Boselli2006, Lisker2007}. Environmental processes affect all galaxies, but especially small dwarf-like galaxies, as can be seen from the morphology-density relation \citep{Dressler1980,Geha2012}. Cluster galaxies, and also those in the field, come in different shapes and sizes and are found to have a large range of masses, from massive elliptical galaxies of around $10^{11}$ M$_\odot$, down to dwarf galaxies with postulated progenitor dark matter halo masses as low as $10^5-10^7$ M$_\odot$ (see \citealt{NaabT2006,BLAND2015,Verbeke2015}). The densest regions in galaxy clusters are also known to be dominated by early-type galaxies \citep{Jerjen2005}. In this study, we focus on a dwarf galaxy in the mass range between $10^8-10^9$ M$_\odot$.

One way to obtain information about the formation of these galaxies, is to examine their stellar populations, since they provide a fossil record of the evolution of the galaxy. This allows the chemical composition and the star formation history (SFH) of the system to be investigated. 
In the last decade, many technological developments (Integral field spectroscopy with higher spatial and spectral resolution, higher throughput optical detectors, etc.) have made it possible to directly study the previously challenging class of dwarf galaxies, as their low surface brightness nature made it extremely difficult to obtain high signal-to-noise spectra of these systems.

Our current knowledge is still rather basic, for example, the unresolved stellar Initial Mass Function (IMF) of dwarf galaxies has not yet been constrained in detail. An attempt to study the normalisation of the IMF has been made by \citet{Tortora2016}, based on a hybrid approach. However, to date, no dwarf galaxy has yet been examined in the same way as massive early-type galaxies (ETGs), i.e through gravity sensitive absorption features, which can be used, in principle, to constrain the dwarf-to-giant ratio (i.e. the slope of the stellar IMF). \citep{VanDokkum2010,Conroy2012b,Ferreras2013,LaBarbera2013,Spiniello2014}. Furthermore, little is known about the chemical abundance ratios in dwarf ellipticals beyond the Local Group. Results have been presented by \citep{Michielsen2003,Hilker2007,Michielsen2007,Michielsen2008r,Chilingarian2009,Koleva2009a,Koleva2011,Paudel2011}.
\subsection{Dwarf elliptical galaxies}
Dwarf elliptical galaxies (dEs) constitute a very important and fascinating class, dominating clusters and massive groups of galaxies in numbers.
By comparing the contribution to the total luminosity function of clustered galaxies with galaxies in the field, it can be seen that dEs are much more abundant in clusters and groups while almost none are found in the field \citep{Dressler1980,Bingelli1988,Lisker2007}. \citet{Geha2012} showed that most quenched dwarf galaxies, i.e red galaxies with little star formation, reside within 2 virial radii of a massive galaxy, and only a few percent beyond 4 virial radii. Galaxy clusters can therefore be viewed as an excellent environment to study the formation and evolution of these systems.

Intrinsic properties and stellar populations of dEs have not yet been studied in as much detail as their giant counterparts. Systematic studies include \citet{VanZee2002,VanZee2004,Lisker2006b,Lisker2007a,Michielsen2008r,Koleva2009a,Janz2014,Rys2015}. dEs are defined to have magnitudes $M_{B_T} \geqslant -18$ \citep{Sandage1984}, where a further distinction between bright and faint dwarfs is made at $M_{B_T}$ around $-16$ \citep{Ferguson1994}. Most observed ellipticals are objects conforming to a S\'{e}rsic surface-brightness profile with exponent between 1 and 1.5 \citep{Caon1993,Koleva2009a}, but there are also compact low-mass early types, commonly known as cEs and similar to the prototype M32, whose S\'{e}rsic indices are usually larger. Such compact ellipticals (cE) are categorised by \citet{Kormendy1985, Kormendy1987,Bender2015} as ellipticals, as in the case of M32 \citep{Ferguson1994}, while he categorises low mass ETGs as spheroidal galaxies (see also \citealt{Guerou2015}).

In general dEs are red, and lie on the well-established correlation of the optical colour and central velocity dispersion, $\sigma$, indicating that the stellar populations are generally old, with sub-solar metallicities \citep{Michielsen2008,Koleva2009a,Janz2009,Paudel2011}. An approach to learn more about the formation episodes that took place in a galaxy is to look at the observed stellar population gradients in the galaxy. Population gradients hold information about the processes, including gas dissipation and merging, that have been playing a role during the formation of the galaxy (see \citealt{White1980,DiMatteo2009,Hirschmann2015}. 
Colour gradients can be used as a tracer for metallicity variations in these systems \citep{DenBrok2011}, while they also claim that more massive galaxies have larger gradients than dwarfs. The observed colour gradient might also have correlations with other structural parameters, such as effective radius, effective surface brightness and, especially, S\'{e}rsic index \citep{DenBrok2011}. This therefore indicates the importance of studying the spacial variations in dwarfs, which are the focus of this paper.

Chemical abundances play an important role. They provide us with information about the stellar populations and yields of all stars in the galaxy \citep{Boehringer2009}. It is known from observational evidence that dEs also harbour multiple stellar populations, which includes younger central regions, with mostly older populations in the outskirts. The central chemical abundances of e.g. Ca and Na proved to be peculiar when compared to abundances found in the local environment and in their more massive elliptical counterparts \citep{Michielsen2003,Zieleniewski2015}. It has also been shown that more massive dEs often have nuclear star clusters \citep{Cote2006}, as is also the case for NGC 1396. 

Measurements of the near infra-red CaT index line strength are found to be very high for dwarf galaxies \citep{Michielsen2003}, in contrast to the lower than predicted CaT values recorded for massive ellipticals (Es) \citep{Saglia2002,Cenarro2003}. These elevated Ca values in dwarfs are also part of the so called Calcium triplet puzzle \citep{Michielsen2003,Michielsen2007}, where it was initially noted that the CaT index \textit{anti-correlates} with the central velocity dispersion ($\sigma$) in bulges of spiral galaxies and in Es \citep{Saglia2002,FalconBarroso2003,Cenarro2003}. This anti-correlation was also shown to be present in the dwarf galaxy regime \citep{Michielsen2003}, where the measured CaT* values were larger than expected for the age-metallicity relation of dwarfs. This CaT-$\sigma$ anti-correlation is in contrast to line indices like Mg$_2$ and Mgb and most other metal lines, which correlate positively with $\sigma$. This is also the case for Ca at lower metallicities \citep{Saglia2002,Tolstoy2009,Grocholski2006}.

In this study of NGC 1396, MUSE data were obtained in order to measure kinematics and obtain information about the stellar populations and chemical abundances of a typical undisturbed early-type dE galaxy and its globular cluster system. This study forms part of a program dedicated to the dwarf galaxy properties in the Fornax core. The globular cluster system is also a good tracer of the original angular momentum from the galaxy's early evolution, and can be used to model the mass distribution in the outer regions of the galaxy \citep{Brodie2015}. Moreover, in order to determine the evolutionary state in which NGC 1396 currently resides, it is necessary to characterise all its properties as a function of galacto-centric radius. These properties include the profiles of dynamical-to-stellar mass and angular momentum obtained from dynamical modelling, the kinematical connection between the field stars and the globular cluster system and the chemical abundance ratio and age/metallicity profile in the galaxy.

In this paper, we investigate the characteristics of the dwarf galaxy NGC 1396 with the focus on its stellar populations. The second paper will include the kinematics of this galaxy. A third paper will focus on the globular cluster system of NGC 1396, and the global mass distribution. This paper is structured as follows: In section 2, we present the general properties of the galaxy under study together with known available photometric data relevant for this study. Information about the observations and the general data reduction procedure is presented in Section 3, together with a description of the data analysis procedure which includes the measurement of line strength indices and the use of population models. Results of the obtained abundance ratios and stellar populations of NGC 1396 will be covered in Section 4, followed by the discussion, conclusions and summary in the last two sections. 
\section{General properties of NGC 1396}
The Fornax cluster is the second largest cluster dominated by early-type galaxies located within 20 Mpc. The cluster is more compact than the larger Virgo cluster, making it a good target for the study of environmental influences on galaxy formation \citep{Jordan2007}.

\begin{figure}
\begin{minipage}[l]{8.5 cm}
\begin{center}
 \includegraphics[width=8.5 cm ,height=5 cm,keepaspectratio=true]{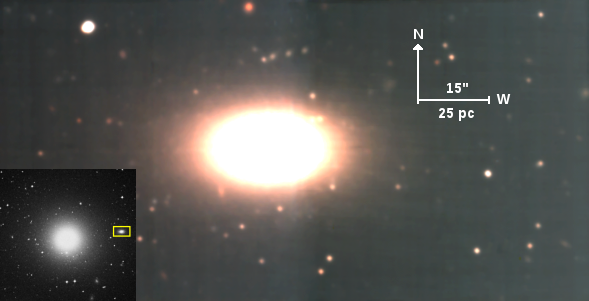}
 \caption{Colour image of NCG 1396 created from the MUSE data set used in this study. The central galaxy in Fornax, NGC 1399, is located at a projected distance of $\sim 5$ arcmin in an ESE direction. The central Fornax region centred on NGC 1399 is indicated with a DSS2 insert in the bottom left corner with the MUSE field of NGC 1396 in the yellow box.}
 \label{ngc1396colour}
 \end{center}  
\end{minipage}
\end{figure}

NGC 1396 (see Figure \ref{ngc1396colour}) is located in close projection to the massive central elliptical galaxy NGC 1399. It has been catalogued in the Fornax cluster catalogue (FCC) as FCC 202 (see also Table \ref{table:parameters}), \citep{Ferguson1989a}, and the galaxy has been observed as part of the ACS Fornax Cluster Survey (ACSFCS, \citealt{Jordan2007}). With a recession velocity of $-600$ km/s from the central cluster galaxy NGC 1399, it is believed to be in a radial orbit at the edge of the cluster escape velocity \citep{Drinkwater2001}. This picture is in agreement with the fact that no isophotal disturbance has been observed for this dwarf galaxy, located at close cluster-centric distance to the central galaxy NGC 1399. However, from the surface brightness fluctuation (SBF) distance measurements made using ACSFCS, NGC 1396 is placed at $20.1 \pm 0.8$ Mpc, while the distance of NGC 1399 is given at $20.9 \pm 0.9$, which agree within the errors \citep{Blakeslee2009}. It is therefore not clear whether NGC 1396 is a foreground object relative to NGC 1399. NGC 1396 has a $V$ magnitude and effective radius of 14.88 and 10.7", respectively \citep{Hilker1999}.  

\begin{table}
\begin{center}
\caption{General properties of NGC~1396}
\begin{tabular}{lcccc}
\hline
\hline
Quantity	& Value	& Unit	& Reference	& \\
\hline
Hubble Type	& d:E6,N			& --			& (1)	& \\
R.A. (J2000)	& $03^{h}38^{m}06.54^{s}$			& HMS	& (2)	& \\
Dec. (J2000)	& $-35^{d}26^{m}24.40^{s}$			& DMS	& (2)	& \\
Helio. Rad. Vel.	&808$\pm$22	& km/s		& (2)	& \\
m$_{V}$			& 14.88	& mag		& (4)	& \\
Distance		& 20.1	& Mpc		& (5)	& \\		
M$_{V}$ [Abs.]			& -16.60 & mag 		& 	& \\
A$_{V}$			& 0.04	& mag		& (3)	& \\
Effective Radius	& 10.7	& arcsec	& (4)	& \\
\hline
\end{tabular}
\label{table:parameters}
\end{center}
\hspace{0cm} {\footnotesize
References:(1) \citet{Ferguson1989a}; (2) \citet{Jordan2007}; (3)\citet{Turner2012}; (4) \citet{Hilker1999}; (5) \citet{Blakeslee2009}}
\end{table}
 
\begin{figure}
\begin{minipage}[l]{8.5 cm}
\begin{center}
    \includegraphics[width=8.5 cm,height=10 cm,keepaspectratio=true]{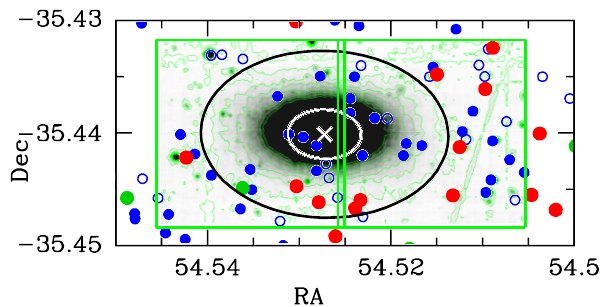}
    \caption{MUSE field pointings for NGC 1396 indicated with two overlapping green squares, with positions (J2000) of globular clusters surrounding the galaxy from the ACS Fornax cluster survey. Blue circles: All GCs selected for the field of NGC 1396 from ACS GC catalogue \citep{Jordan2009}; blue filled circles: Bona-fide GCs with a probability P(GS) $\geq$ 0.5 (see selection criteria by \citealt{Jordan2009}); red and green circles: Radial velocity confirmed GCs with $500<$ v $<2500$ km/s (\citealt{Schuberth2010},Pota et al., in prep); green circles: GCs with radial velocities within 150 km/s from the radial velocity of NGC 1396 (NED: $808 \pm 150$ km/s). The white and black ellipses indicate the effective radius and the radius corresponding to D25 (B surface brightness=25 mag/arcsec$^2$) of the system, respectively.}
    \label{muse_pointings}
  \end{center}  
\end{minipage}
\end{figure}
From photometric analysis, NGC 1396 also contains a distinct nuclear star cluster (NSC) component. With an effective (half-light) radius in the g- and z- bands of $0.047 \pm 0.003$, this NSC accounts for a $\sim1\%$ contribution to the luminosity profile of the galaxy outside of 0.5 arcsec \citep{Turner2012}. It is therefore not possible to study the properties of the NSC using ground based data, without adaptive optics assistance. Our data have typical seeing values of $\sim 1$ arcsec, thus the NSC remains inaccessible for detailed analysis. Apart from this, indications of a globular cluster (GC) system is also found around NGC 1396, where about 40 candidate GCs have been identified in the ACS survey, within a radius of $\sim4.8$ arcmin \citep{Jordan2009}. A few of these are already confirmed to belong to NGC 1396 with radial velocity measurements, as it is shown, together with the MUSE field pointings, in Figure \ref{muse_pointings} \citep{D'Abrusco2016}.
\section{Data and Analysis}
\subsection{Observations}
\begin{figure}
\begin{minipage}[l]{8.5 cm}
 \begin{center}
 \includegraphics[width=8.5 cm ,height=10 cm,keepaspectratio=true]{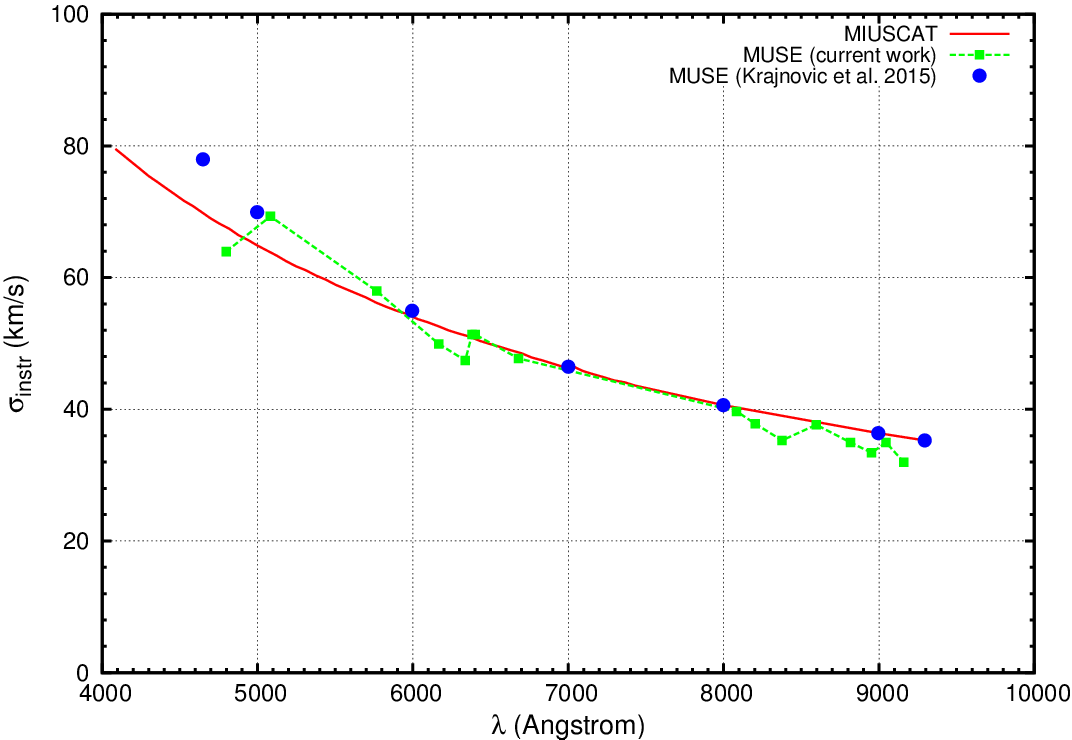}
 \caption{MUSE spectral resolution, as measured from our data (green curve), as a function of wavelength. Notice the good agreement with estimates of MUSE instrumental resolution from \citet{Krajnovic2015};(blue dots). The MUSE resolution matches almost exactly (being slightly higher in the blue region) that of the MIUSCAT stellar population models (FWHM $\sim2.51$ \AA{}) \citep{Vazdekis2010}, used to analyse the MUSE spectra of NGC 1396 in this paper. The MILES resolution was also measured to be $2.51$ \AA{} \citep{FalconBarroso2011}}
 \label{resolution}
 \end{center}
\end{minipage} 
\end{figure}

We used the new spectrograph at ESO, MUSE. The MUSE instrument was commissioned on the VLT telescope with first light on January 31st 2014. It consists of 24 IFU modules with a total field of view of $1 \times 1$ arcmin in the wide field mode, covering a broad wavelength range of 4750-9350 \AA{}. The MUSE data have a spatial sampling of $0.2$ $\times$ $0.2$ arcsec. 
The measured instrumental resolution ($\sigma_{inst}$) as a function of wavelength, measured from features in a sky flat field exposure, is shown in Figure \ref{resolution}, and spans a range between 35 km/s at 9300 \AA{} and 65 km/s at 4650 \AA{}.

NGC 1396 was observed during the 4 nights of December 15 2014, January 17,19 and 20, 2015, with average seeing values ranging between 0.9 and 1.5 arcsec. The galaxy was observed using two pointings, East and West, with an overlap of ~4 arcsec, in order to cover both sides of NGC 1396 together with a large portion of the GC systems belonging to this galaxy. In total eight hours were awarded to this project (4 hours per side with a total of 16 exposures). The 16 exposures were arranged in observing blocks that consisted of two exposures of 1300 s on the galaxy followed by one offset sky exposure of 70 s for the later use in sky subtraction. The exposures were also taken with a dither pattern that involves a change in position angle of $90^\circ$. The total exposure time on target therefore ads up to $\sim5.9$ hours excluding overhead time. The configuration, as seen in Figure \ref{muse_pointings}, was chosen to enable the study of all components of the system out to a radius of 60 arcsec ($\sim6$ kpc at distance of Fornax) along the major axis direction. Together with the MUSE pointings, Figure \ref{muse_pointings} shows positions of GCs, where blue filled circles indicate Bona-fide GCs with a probability of being a GC, greater that 50\% \citep{Jordan2009}. Red and green circles show all confirmed GCs within $500<$ v $<2500$ km/s where the green circles denote the subset with radial velocities within 150 km/s of that of NGC 1396, with heliocentric velocity of $808\pm22$ km/s. The effective radius (small ellipse) and the radius corresponding to D25 (large ellipse, B surface brightness=25 mag/arcsec$^2$) are also indicated in Figure \ref{muse_pointings}.  

\begin{figure*}
\begin{minipage}[l]{\textwidth}
\begin{center}
    \includegraphics[width=18 cm,height=10 cm,keepaspectratio=true]{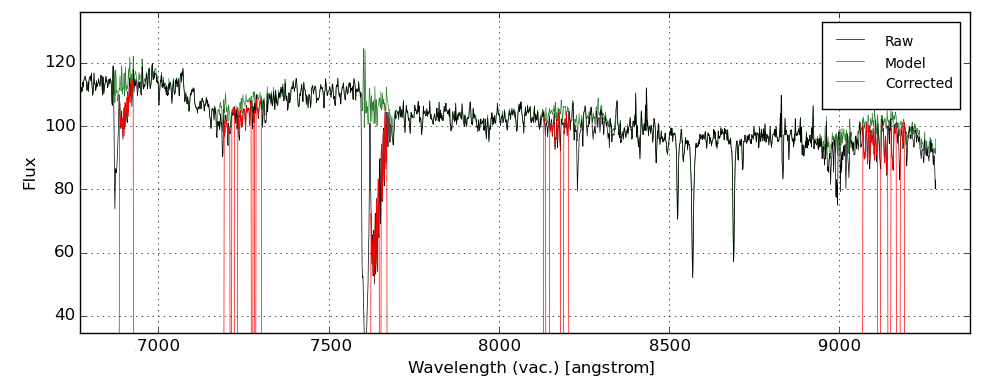}
    \caption{Telluric atmospheric transmission model (red) created with the MOLECFIT routine, applied to the full extracted central spectrum (black), showing the telluric corrected spectrum in green. The regions where the model (red) is over-plotted to the observed spectrum (black) are the ones used in the fitting with MOLECFIT. The central observed spectrum was extracted using a square aperture of $15 \times 15$ spaxels.}
    \label{telluric}
  \end{center}  
\end{minipage}
\end{figure*}
\subsection{Reduction}
Basic data reduction and pre-processing of the raw data were done using the ESOREX standard MUSE pipeline (version 1.0.0), which deals with the basic reduction processes of bias subtraction, flat field correction, wavelength calibration, flux calibration, sky subtraction, error propagation together with other distinctive IFU spectroscopy features i.e. illumination correction and re-sampling of the final reconstructed cube. All of these processes are performed on data in the form of pixel tables to avoid intermediate re-sampling steps. The pixel table format is used to store the non-resampled pixel value together with the corresponding coordinates, data quality and statistical error. The following steps were performed on the raw data in order to produce the fully reduced data cube:
\begin{itemize}
\item The basic master calibration frames which include bias, arcs and flat fields were created as specified in the ESO MUSE Pipeline User Manual. The observed flat fields were combined into an illumination correction cube which is used in successive pre-processing steps. 
\item Wavelength calibration was done by using a set of arc lamp exposures, consisting of Ne, Xe, and HgCd, to compute the wavelength solution. 
\item The master bias frame were subtracted from each science exposure which includes the sky fields and standard star observations.
\item No dark frames were necessary due to a very small dark current in the MUSE CCDs.
\item Flat fielding and flux calibration were applied to each of the pixel tables. Sky frames were created from a set of dedicated sky exposures. The sky frames were calibrated, combined and subtracted during the main reduction routine.
\item Offsets between different pointing positions were calculated, using catalogue coordinates from the FDS survey (Fornax Deep Survey:PI. R.F. Peletier and M. Capaccioli). This was done for each exposure using 2D Gaussian fitting on six different point like objects in the field. Corrections were applied to the table header files before combination.

\item During the final combination phase, in which 16 pixel tables were combined simultaneously, an inspection of the extracted central spectrum of the final data cube, showed an unexplained step in the continuum around 7000 \AA{}.  This proved to be a problem occurring only when combining multiple pixel tables with significant offsets from one another in RA and DEC.

Due to the afore mentioned complications experienced in the combination process, the reduced pixel tables were combined in groups of four (closest in observation time). Each group consists of sets of observations from both sides of the galaxy and were re sampled into a cube. This resulted in four fully reduced cubes, each re-sampled to a spectral sampling of 1.25 \AA{}/pix. The four cubes were aligned, trimmed to a common section, and combined, into a single data cube for further analysis. The combination was done by using a weighting scheme, in which weights were assigned to each of the four cubes, depending on their data quality and S/N ratio.
Because of the variation of the sky background with time, sky residuals still had to be removed. This second sky subtraction procedure was done by assuming that the contribution of the galaxy was negligible in the outer parts of the mosaic. For every horizontal line in each wavelength frame of the resampled cube, the sky was determined by taking the median of two horizontal bands across the image of each frame, North and South of the galaxy, and subtracting this from each wavelength frame of the data cube. Additional uncertainties have also been added to the error budget, in which the flux errors also account for uncertainties in the telluric correction (see below). On top of that, a further component has been added to the statistical errors of the line strengths in order to account for uncertainties in sky subtraction, as well as uncertainties related to binning and combining of data cubes.
\end{itemize}

\subsubsection{Telluric correction}
Telluric correction was disabled in the ESOREX standard MUSE pipeline, and was done instead with the standalone software MOLECFIT (\citealt{Smette2015} and \citealt{Kausch2015}). MOLECFIT is used to derive a correction function for the removal of telluric features by taking an atmospheric profile together with a set of pre-selected molecules as input to the radiative transfer code. 

A median combined central spectrum ($15 \times 15$ spaxels) of the reduced cube was extracted from which the telluric correction was computed. The fit by MOLECFIT was limited to regions where telluric lines are prominent, as marked in Figure \ref{telluric}.  With an iterative sequence, MOLECFIT varies the atmospheric profile in order to create a model that fits the science spectrum. For the fit to be successful in these regions, true absorption lines were first identified from a MIUSCAT \citep{Vazdekis2012} stellar population model (with Kroupa IMF, age of 6 Gyr and [Z/H]=$-0.4$ ; i.e. suitable values for NGC 1396, see Section\ref{SPA}) and masked out in the fitting routine. The input parameters of MOLECFIT were also varied, where it was found that the output telluric model is consistent at a level of 1\% throughout the whole wavelength range. The resulting telluric correction was applied to the full wavelength range, as shown in Figure \ref{telluric}, which shows the observed spectrum (black), the telluric atmospheric model (red), and the telluric corrected spectrum (green). The telluric correction was applied by dividing each spectral pixel in the data cube by the atmospheric transmission model, after which further analysis steps were done on the reduced data cube. 

\subsection{Extracting kinematical information}
\begin{figure}
\begin{minipage}[l]{8 cm}
\begin{center}
    \includegraphics[width=8.5 cm,height=9 cm,keepaspectratio=false]{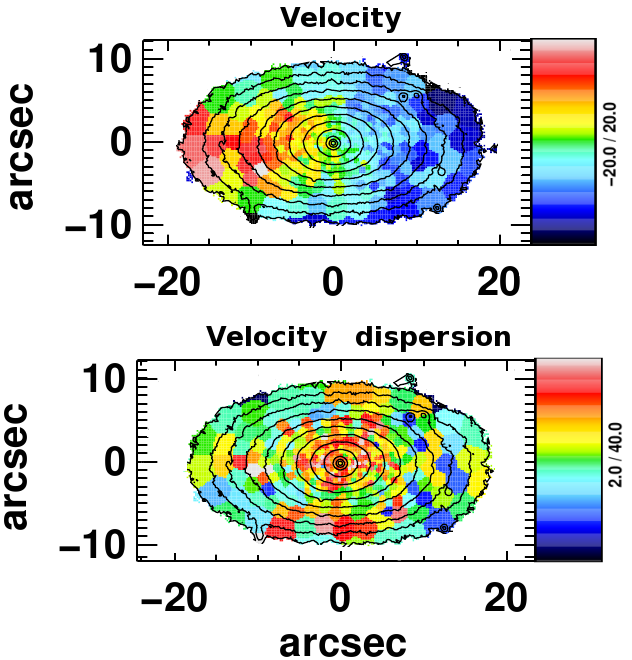}
    \caption{Velocity and velocity dispersion fields of NGC 1396, with a measured mean velocity and velocity dispersion of $\sim840$ km/s and $\sim25$ km/s respectively (measured in the red part of the spectrum) (see Mentz et al. (2016) in prep).}
    \label{kinematics}
  \end{center}  
\end{minipage}
\end{figure}

To perform the stellar population analysis, each spaxel in the reduced data cube was first corrected for its (position-dependent) recession velocity. 
The Voronoi tessellation binning method from \citet{CAPPELLARI2003} was used to bin the IFU data to a constant S/N ratio of 100 per bin. 
Kinematic information is extracted from the Voronoi binned data by fitting the binned galaxy spectra using the penalized pixel fitting routine (pPXF), developed by \citet{CAPPELLARI2004}. This routine makes use of template stellar spectra to fit the absorption line spectrum of the galaxy in order to find the best fitting kinematics. The velocity and velocity dispersion fields are shown in Figure \ref{kinematics}.

To run pPXF, a total of 63 SSP models were used as templates, from the MIUSCAT library \citep{Vazdekis2012}, with metallicity ([Z/H]) ranging from $-2.32$ to $+0.22$, and age between $0.1$ to $17.8$ Gyr. Before making any radial index measurements, pPXF was run on each Voronoi binned spectrum, deriving its recession velocity, v. Then a velocity v was assigned to each spaxel in the data cube by cross-correlating its position to pixels in the parent Voronoi bins.

\subsection{Extracting absorption indices}
\label{radbins}
\begin{figure}
\begin{minipage}[l]{8.5 cm}
\begin{center}
    \includegraphics[width=8.5 cm,height=10 cm,keepaspectratio=true]{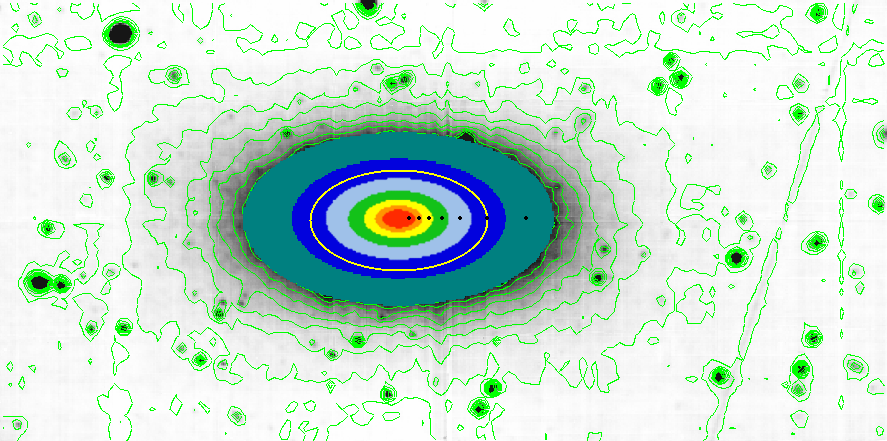}
    \caption{Contour map of NGC 1396 with radial bins overlaid on NGC 1396, with luminosity-weighted bin centres indicated with black dots. The effective radius (R$_e$=10.7 arcsec) is indicated with a yellow ellipse.}
    \label{contour_radbin}
  \end{center}  
\end{minipage}
\end{figure}

In this paper we focus on the information obtained on stellar populations by measuring different absorption features, which also include features in the near infrared.
To study the radial variation of spectral indices in NGC 1396, the data were binned in elliptical annuli in order to be able to measure indices at different radial distances from the centre of the galaxy. In order to best fit the galaxy, the elliptical bins, with a major- to minor axis ratio of 2 and PA of 90 degrees, were constructed to have a width of $(B\times f)^n$, where B was chosen to be around the typical seeing (FWHM) value during the observing night, and $f$ a parameter introduced to enable varying the width of the radial bins. The binning scheme can be seen in Figure \ref{contour_radbin}, where the luminosity-weighted radii are indicated with black dots on the radial bins and the effective radius is shown by the yellow ellipse.   
Due to the variable resolution between the blue and red part of MUSE, we also measured the $\sigma$ of each radially binned spectrum by running the software STARLIGHT in the blue (4700-5700 \AA{}) and red (8000-8730 \AA{}) spectral regions, feeding STARLIGHT with MIUSCAT SSP models (see Section \ref{starlight}). This $\sigma$ provided us with the amount of broadening necessary to match the resolution of the MIUSCAT models to our data in the blue and red spectral regions, respectively (see also Figure \ref{resolution}). We found broadening values for the blue region between 45 and 51 km/s and for the red region between 26 and 31 km/s, with median values (among all spectra) of 49 and 28 km/s, respectively, and no significant radial trend of $\sigma$. 

For the stellar population analysis, Lick indices were measured at the nominal resolution of each radial binned spectrum after which it was compared to MIUSCAT models. This was done in order to avoid any smoothing of the data, while maximising the information to be extracted. Lick indices were measured on the radial binned spectra by using the index task from the \reduceme\ package by N. Cardiel \citep{Cardiel2015}.

\begin{figure*}
 \centering
 \includegraphics[width=15 cm ,height=10 cm,keepaspectratio=true]{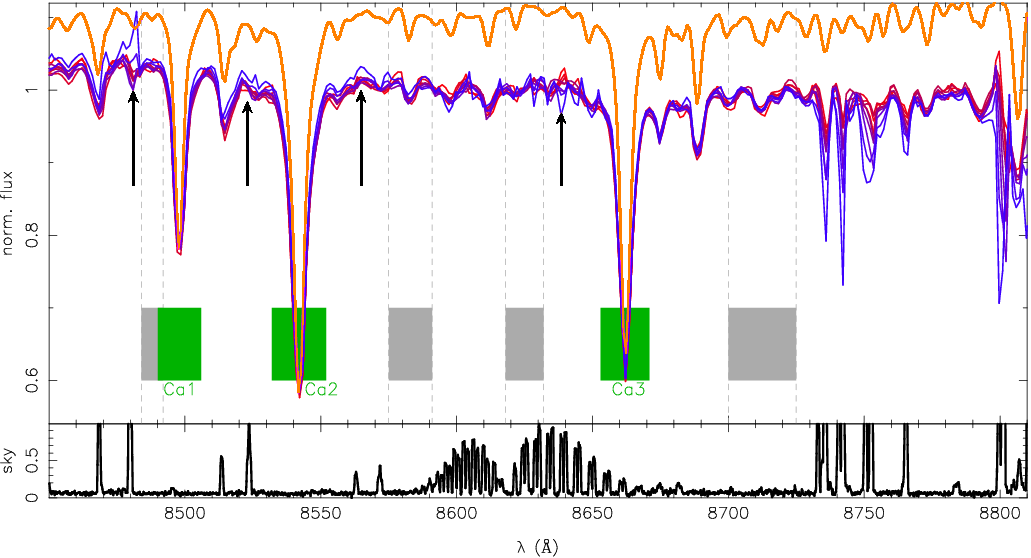}
 \caption{Spectral region of the CaT where all binned spectra (colour coded from the innermost bin in red to the outermost bin in blue) are over-plotted together with a MIUSCAT model ([Z/H]=$-0.4$ and age $\sim5.6$ Gyr) in orange as reference. Black arrows show regions with potential sky residuals. The optimised definition for the CaT pseudocontinua is shown in grey for the exclusion of regions of strong sky residuals while the passband definition (green) is the same as in \citet{Armandroff1988}. A sky spectrum can be seen in the bottom panel, shown in black.}
 \label{CaTdef}
 \end{figure*}

In order to minimise the effect of the sky residuals, present in the blue continuum band of the CaI index (especially in our outermost spectra), different definitions of the pseudo-continua were adopted for this index, while keeping the \citet{Armandroff1988} central passband definition. This is tabulated in Table \ref{IndexdefT} together with some of the other index bands of importance in this study. In Figure \ref{CaTdef} we show a plot of all radially binned spectra over-plotted (colour coded from the inner bin in red to the outer bin in blue) in the region of the CaII IR triplet, where the sky residuals (black arrows) in the spectrum at 8481 \AA{} are clearly visible together with the corresponding sky emission line in the lower panel of the figure. This adjusted definition of the pseudocontinua was done while also making sure that all three Ca lines have a similar sensitivity to age and metallicity, as well as single elemental abundances (based on the public available version of Conroy \& van Dokkum (CvD) stellar population models) as in the \citet{Cenarro2001} definition. The same index definition was used both for indices and models. 
\newline
\begin{table*}
\begin{center}
\caption{Index band definitions}
\begin{tabular}{lccccc}
\hline
\hline
Index & Blue bandpass \AA{}	& Central bandpass \AA{}	& Red bandpass \AA{}	& Source& \\
\hline
H$\beta_{o}$&4821.175-4838.404&4839.275-4877.097&4897.445-4915.845&(1)& \\
Mg$_{b}$&5142.625-5161.375&5160.125-5192.625&5191.375-5206.375& (2)& \\
Na5895&5860.625-5875.625&5876.875-5909.375&5922.125-5948.125&(3)& \\
TiO1&5816.625-5849.125&5936.625-5994.125&6038.625-6103.625&(4)& \\
TiO2&6066.625-6141.625&6189.625-6272.125&6372.625-6415.125&(5)& \\
NaI8190&8143.000-8153.000&8180.000-8200.000&8233.000-8244.000&(3b)& \\
Ca1&8484.000-8492.000&8490.000-8506.000&8575.000-8591.000&(6)& \\
Ca2&8484.000-8492.000&8532.000-8552.000&8575.000-8591.000&(7)& \\
Ca3&8618.000-8632.000&8653.000-8671.000&8700.000-8725.000&(8)& \\
\hline
\end{tabular}
\label{IndexdefT}
\end{center}
\hspace{0cm} {\footnotesize
References:(1) \citet{Cervantes2008}; (2),(3),(4),(5) Lick definition; (3b) \citet{LaBarbera2013}; (6) Optimised; (7) \citet{Armandroff1988}; (8) Optimised}
\end{table*}
 
\subsection{Stellar population analysis}
\label{SPA}
Stellar population parameters i.e. age, metallicity ([Z/H]), IMF, and Ca, Mg, and Na abundance ratios were determined by means of i) full-spectral fitting and ii) by comparing the measured Lick indices from the elliptically-binned spectra with those calculated from SSP models with varying age, metallicity, and IMF, accounting for the effect of varying abundance ratios. With the purpose of testing the robustness of the derived parameters, results were compared between the two full-spectral fitting tools, STARLIGHT \citep{STARLIGHT} of the SEAGal (Semi Empirical Analysis of Galaxies) Group and pPXF, and with parameters derived by a simultaneous fit of different spectral indices. It should also be noted that the underlying methods used in STARLIGHT and pPXF differ quite significantly, in terms of the starting assumptions that effectively linearise the complex physical processes in SEDs, as is done in pPXF, compared to STARLIGHT which does not need the linearity condition \citep{Walcher2011}. 
In the following subsection, different techniques are described that were applied to extract stellar population parameters from the binned spectra. 
\begin{figure}
 \begin{minipage}[l]{8.5 cm}
  \begin{center}
  \includegraphics[scale=0.2]{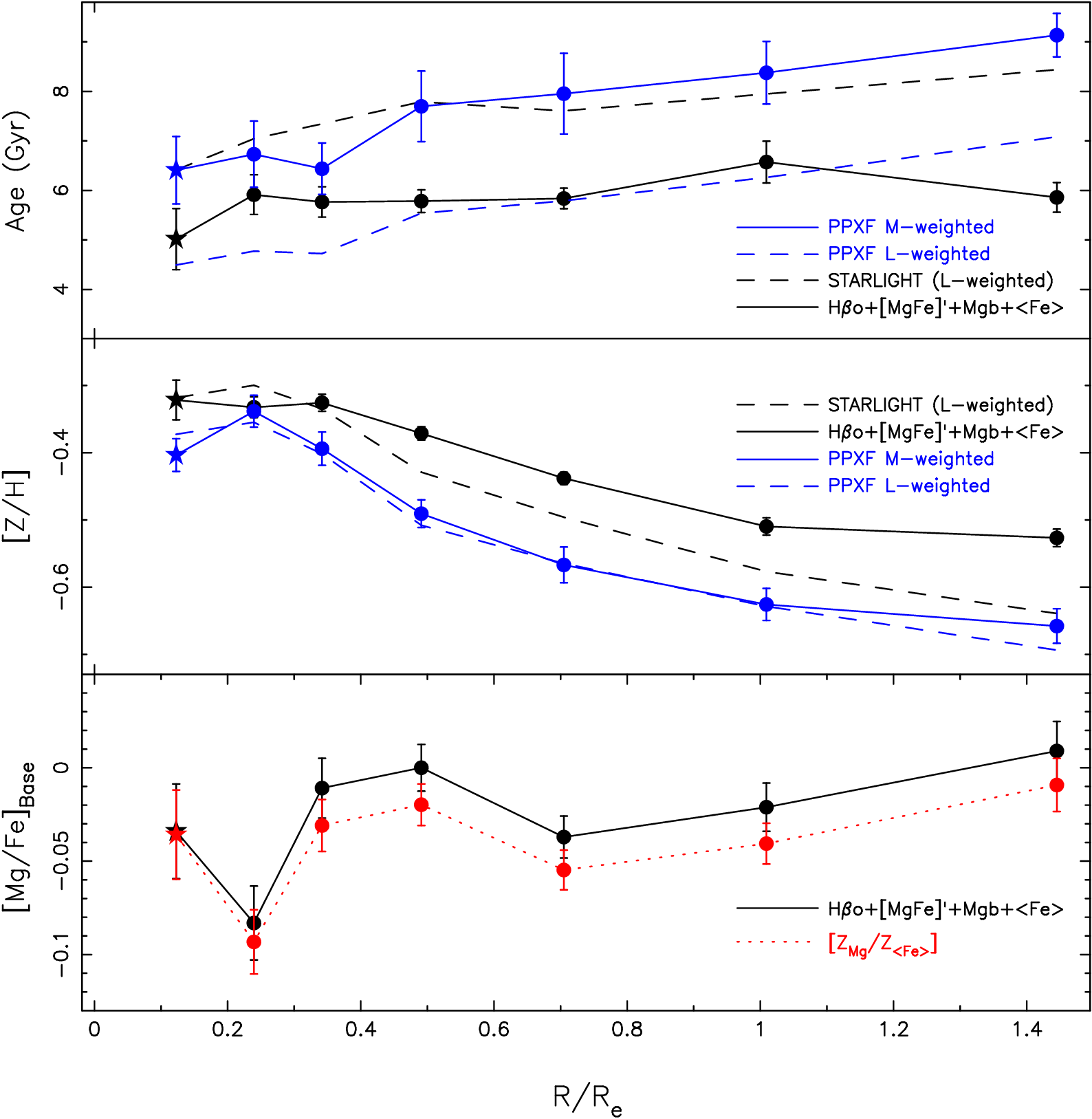}
 \caption{Age, Z, and [Mg/Fe] profiles as a function of galacto-centric radius obtained from different methods by using STARLIGHT, pPXF and index fitting. Results from STARLIGHT and pPXF are presented as black and blue lines respectively, index fitting is indicated by filled black circles, and the [$\mathrm{Z_{Mg}/Z_{Fe}}]_{Base}$ values are shown by the red line. The filled star in each panel indicates the NSC-contaminated bin.}
 \label{AgeZalpha}
 \end{center}
 \end{minipage}
\end{figure}
\subsubsection{STARLIGHT}
\label{starlight}
With STARLIGHT, a base of 135 SSP templates were used, which were constructed from MIUSCAT \citep{MIUSCAT} models, ranging in metallicity and age from $-1.31$ to $0.22$ and from $0.5$ to $17.78$ Gyr, respectively, and assuming a Kroupa IMF. The MIUSCAT models provide us with extended coverage compared to the MILES \citep{SanchezBlazquez2006} models with the addition of the CaT region (8350-9020\AA{}) \citep{Cenarro2001}, and the region covered by the original INDO-US \citep{Valdes2004} spectra, required for analysing features in the broad MUSE wavelength range. With these models as input, we derive, using STARLIGHT, the best-fitting linear combination of the SSPs, obtaining the broadening $\sigma$ (see Section \ref{radbins}), and the extinction $A_v$. From the STARLIGHT best-fitting SSP mixture, the luminosity-weighted age and metallicity as a function of galacto-centric distance were computed. For each spectrum, the [Mg/Fe] abundance ratio was computed using the same approach as in \citet{LaBarbera2013}, hereafter LB13. At fixed (luminosity-weighted) age, we computed the difference between metallicity from Mgb and $<$Fe$>$ lines, [$\mathrm{Z_{Mg}/Z_{Fe}}$], and converted it into [Mg/Fe] ($\sim\mathrm{Z_{Mg}/Z_{Fe}}$; see LB13 for details). Notice that the computation of [$\mathrm{Z_{Mg}/Z_{Fe}}$] is performed with MIUSCAT stellar population models, which rely, in the optical range ($<~7400$Ang), on MILES stellar spectra. These spectra follow the characteristic abundance pattern imprinted by the SFH of our Galaxy. Since the models are computed with scaled-solar isochrones and assuming [Z/H] =[Fe/H] for stellar spectra (see \citealt{Vazdekis15models}), the models can be actually regarded as "base" models, and the inferred abundance ratios should be regarded as relative values to the abundance pattern of stars in our Galaxy. For this reason, we refer to [Mg/Fe] estimates in Figure \ref{AgeZalpha} as [Mg/Fe]$_{Base}$, and to other abundance ratios computed with base models, as [X/Fe]$_{Base}$, with X={Mg, Ca, Na} (see below). In what follows, abundances will be labelled with a BASE subscript for abundances derived from BASE-models and without subscript for the MW-corrected abundances. In Figure \ref{AgeZalpha} the results from STARLIGHT are presented as black dashed lines, while [$\mathrm{Z_{Mg}/Z_{Fe}}]_{Base}$ values are shown by the red line. 

\subsubsection{pPXF}
In order to test the robustness of radial trends of stellar population properties extracted with STARLIGHT, we ran pPXF with MIUSCAT models, covering the same range in age and metallicity as STARLIGHT. With the fit of each radial binned spectrum, a $2^{nd}$ degree multiplicative polynomial was used to correct the shape of the continuum. Parameters obtained by using pPXF are shown in Figure \ref{AgeZalpha} as blue lines. With pPXF, both mass- and luminosity-weighted age and [Z/H] were computed, which are represented in Figure \ref{AgeZalpha} as solid and dashed blue lines respectively. As for STARLIGHT, pPXF fits were found to provide a very good description of the data, with residuals at percent level.
\subsubsection{Absorption line indices}
\label{sec:lineindices}
\begin{figure*}
\begin{minipage}[c]{\textwidth}
\begin{center}
    \includegraphics[scale=0.35]{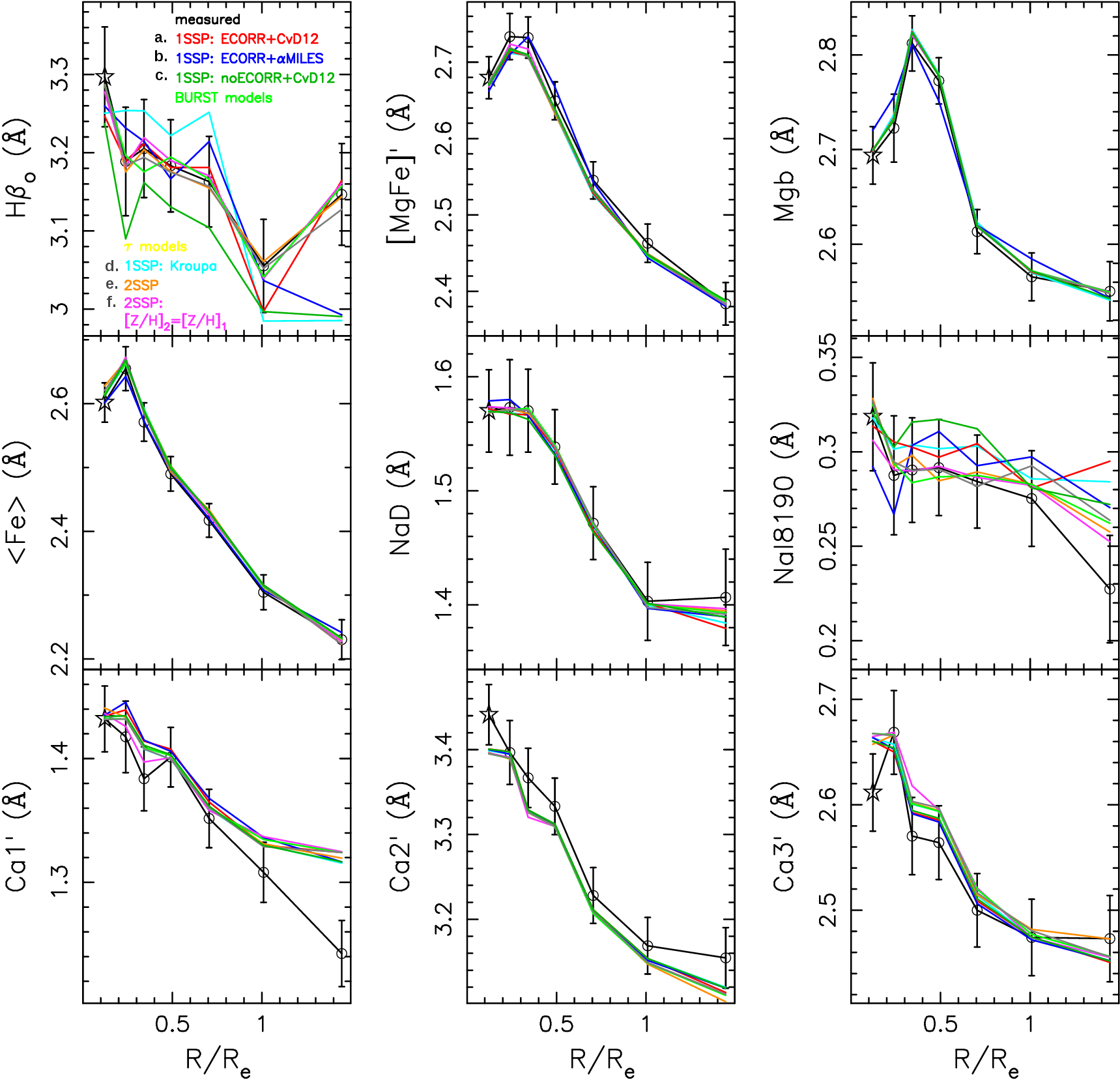}
    \caption{Best-fitting line-strength profiles for the one SSP, two SSP, Burst and Tau model approaches. Measured indices are shown in black with the fits over-plotted. a): One SSP corrected for H$\beta_o$ emission and abundance ratios modelled with CvD12 models, shown in red, b): Same as in a) but with optical indices (H$\beta_o$, NaD, $<$Fe$>$, Mgb) to [Mg/Fe]$_{Base}$) modelled with $\alpha$-MILES models (Vazdekis et al. 2015), using SSPs with Age=5 Gyr, [Z/H]=$-0.35$, and a Kroupa IMF, shown in blue, c): Same as in a) but with no emission correction to H$\beta_o$, shown in green, d): same as in a) but with the IMF fixed to a Kroupa-like distribution ($\Gamma=1.3$), shown in cyan. e) Two SSP models with varying the metallicities of the two components independently, shown in orange. f) Two SSP approach with varying the ages and relative mass fractions of the two components, while keeping the metallicity the same, shown in pink. Additional Burst and Tau models were applied where the Burst models in light green are defined to have constant SFR, and Tau models in grey with an exponentially declining SFR.}
    \label{radindex}
  \end{center}  
\end{minipage}
\end{figure*}

Absorption line indices provide us with an independent method for the estimation of stellar population properties, with the advantage to use selective features in the spectrum, to reduce the effect of degeneracies. We have considered two cases: (i) Fitting  a restricted number of spectral features, in order to estimate only age, metallicity, and  [Mg/Fe]$_{Base}$, and (ii) a more general case, where a larger set of spectral features is fitted, in order to estimate also [Ca/Fe]$_{Base}$ and [Na/Fe]$_{Base}$ (see below). At first (case i), we have fitted Hbo, [MgFe]$_{Base}$, $\langle$Fe$\rangle$, and Mgb, with MIUSCAT model predictions, with a Kroupa IMF, age and metallicity in the range from $0.5$ to $17.78$ and $-1.31$ to $0.22$, respectively. To this aim, we modelled the sensitivity of Mgb5177 and $\langle$Fe$\rangle$ to [Mg/Fe]$_{Base}$ with the publicly available CvD models (for an old age of 13 Gyr, solar metallicity and a Chabrier IMF). The responses of Mgb and $\langle$Fe$\rangle$ to [Mg/Fe]$_{Base}$ are computed at fixed total metallicity, following a similar approach as in \citet{Vazdekis15models}. From this approach, the best fitting parameters, age, [Z/H], and [Mg/Fe]$_{Base}$ were derived, which are shown with black filled circles in Figure \ref{AgeZalpha}.
A fairly good agreement was found between the solar scaled [Mg/Fe]$_{Base}$ from STARLIGHT with the estimates obtained by fitting the above indices simultaneously, as shown in the bottom panel of Figure \ref{AgeZalpha}. 
We also consider the case (ii) where all indices i.e. H$\beta_o$, [MgFe], Mgb5177, $\langle$Fe$\rangle$, NaD, NaI8190, Ca1, Ca2 and Ca3 are fitted simultaneously. In this approach, we consider two cases, where the effect of [Mg/Fe]$_{Base}$ is modelled either with alpha-MILES \citep{Vazdekis15models} or CvD12 models \citep{CvD12}. The effect of other specific abundances, i.e. [Ca/Fe]$_{Base}$ and [Na/Fe]$_{Base}$, is always modelled with CvD models. Notice that in this approach, we have only considered the sensitivity of the selected spectral features to the abundance ratios of their leading elements (i.e. Ca, Mg and Na), as other elements are expected to have a minor role.
The indices TiO1 and TiO2 are not included in the fit due to the fact that both indices are affected by telluric absorption and flux calibration uncertainties together with some residuals from sky emission, which especially affected TiO2 and made it impossible to calculate reliable TiO and Ti abundance ratios. 
The spectra were also fitted with one and two SSP models together with Burst and Tau SFH models to see whether a multi-component population could better fit the observed spectra. Notice that when fitting all indices (case ii), we considered models with a varying IMF (unless stated otherwise). In particular, we consider predictions for MIUSCAT models with a bimodal IMF (\citealt{Vazdekis1996,Vazdekis1997}), having slope $\Gamma_b$ in the range from 0.3 to 3.3. The cases with $\Gamma_b=0.3$ and 3.3, correspond to a top- and bottom-heavy distributions, respectively, while for $\Gamma_b=1.3$, the bimodal IMF approximately matches the Kroupa distribution (see e.g. \citet{Ferreras2015}).
\newline
\newline
\textbf{One SSP:}
\newline
In the case of the single SSP, four different scenarios were tested on the observed spectra as shown in Figure \ref{radindex}. In the first scenario, \textbf{a.)}, Abundance ratios are modelled with CvD12 models while H$\beta_o$ is corrected for emission as explained in \citet{LaBarbera2013}. For this case, the best fitting results are shown as red curves in Figure \ref{radindex}. In the second case, \textbf{b.)}, the response of the optical indices H$\beta_o$, NaD, $\langle$Fe$\rangle$ and Mgb to [Mg/Fe] are modelled with $\alpha$-MILES models (see blue curves in Figure \ref{radindex}. This was done with a single SSP with age and metallicity of 5 Gyr and $-0.35$ dex, respectively, while using a Kroupa IMF. In the third scenario, \textbf{c.)}, the same was done as in \textbf{a.)} but without any emission correction to H$\beta_o$, where this case is shown in green. The last scenario, \textbf{d.)}, were treated similar to scenario \textbf{a.)} but with the IMF fixed to a Kroupa $\Gamma_{b}=1.3$ distribution, which is shown in cyan. 
\newline
\newline
\textbf{Two SSPs:}
\newline
Two different scenarios were used to fit the indices with two SSPs. In the first method (shown in orange in Figure \ref{radindex}(e)) the metallicities of the two components were varied independently as well as their age and relative mass fraction. In the second method (f), (shown in pink), the age of the two components and the relative mass fraction were allowed to vary while the metallicity was assumed to be the same for both components.

Burst and Tau SFH models were also implemented. Burst models are defined to have a constant SFR, with a formation time $t_0$ and star formation time scale of $dt$, while tau models have exponentially declining SFR given by $(exp^{-t/\tau})$, with formation age $t_0$ and scale $\tau$. For comparison these two models were also included in Figure \ref{radindex} as green (Burst model) and grey (Tau model).

\subsubsection{Abundance analysis}
\label{ab_correction}
As stated above (see Section \ref{starlight}), all abundances are obtained using BASE-models, so that the abundance estimates should be interpreted relative to those stars used in the MILES library at a specific metallicity. The stars used in MILES follow the abundance pattern of those in the disk of our Galaxy,  which means that at sub-solar metallicities, these stars have non-solar abundances \cite{Venn2004a}. In order to correct the abundances obtained from BASE-models to solar-scale, for [Mg/Fe], we used the estimates of [Mg/Fe] for MILES stars to derive a relation between [Mg/Fe]$_{Base}$ and metallicity. This relation was used to correct [Mg/Fe]$_{Base}$ into "true" [Mg/Fe] ratios. For the other abundance ratios, we took estimates of [Ca/Fe], and [Na/Fe] for disk stars in our Galaxy (as those estimates are not available for MILES stars), and derived a linear relation between [Ca/Fe] and [Mg/Fe], as well as between [Na/Fe] and [Mg/Fe]. These relations were used to correct [Ca/Fe]$_{Base}$ and [Na/Fe]$_{Base}$ to the "true" abundance ratios [Ca/Fe] and [Na/Fe]. This was done for the range of metallicities ($-0.6$ to $-0.3$ dex) as measured for NGC 1396. These relations were then applied to transform [Mg/Fe]$_{MILES}$ into [Ca/Fe] and [Mg/Fe].       
\begin{figure*}
 \begin{center}
 \includegraphics[width=15 cm ,height=10 cm,keepaspectratio=true]{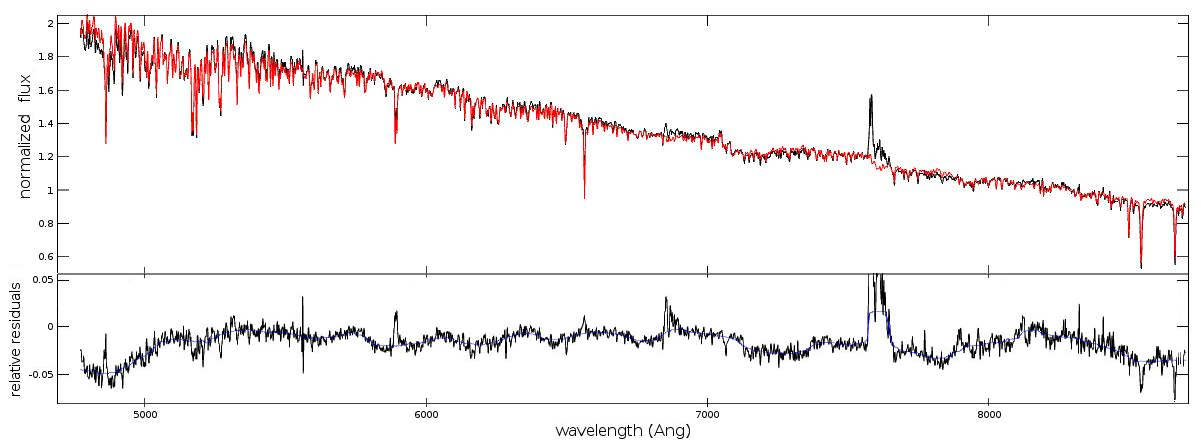}
 \caption{Example of the STARLIGHT fit to the spectrum, extracted from the first elliptical bin. The observed spectrum is shown in black with the model fit over-plotted in red. The bottom panel shows the residual obtained throughout the spectrum with the median smoothed trend in blue.}
 \label{fit}
 \end{center}
\end{figure*}
\newline
\newline
\subsubsection{IMF}
Information about the IMF could be obtained by using a number of different methods (\citealt{VanDokkum2012,LaBarbera2013,Spiniello2014,LaBarbera2016}. In the current paper, a variety of gravity sensitive stellar absorption features from dwarf and giant stars are studied. Prominent features of dwarf stars are the NaI doublet ($\lambda$8183, $\lambda$8195) and the FeH Wing-Ford band ($\lambda$9916), where the latter is beyond the MUSE spectral coverage. For the indices dominated by giant stars, the CaII triplet is the strongest gravity-sensitive feature \citep{VanDokkum2012,Cenarro2003}. Other important IMF sensitive features included in the MUSE spectral range are NaD, TiO1 and TiO2, of which TiO and Na could therefore prove to be very important in constraining the IMF in low metallicity environments as found in NGC 1396 \citep{Spiniello2012, Spiniello2014, Navarro2015}. However, as stated above, TiO features could not be included in the present analysis.
\subsubsection{Gradients}
Radial gradients shown in Table \ref{gradients} were extracted from the different line indices and stellar population parameters, measured form the radial-binned spectra. These parameters were measured from each radial binned spectrum and fitted by using a damped least-squares method, with the exclusion of the central radial bin in order to avoid the effect that a central nuclear star cluster might have on the measured gradients (see e.g. \citealt{DenBrok2011}). The radial extent over which these gradients were measured, covers the region from 2.5 arcsec (0.24 R$_e$) from the centre of the galaxy to a distance of 15.5 arcsec (1.45 R$_e$).

\section{Results}
From the spectral fitting, shown for the full spectrum in Figure \ref{fit}, a good fit was obtained to the observed spectrum (in black), where a good overall fit (over-plotted in red) is obtained to all radial binned spectra, with residuals less than $\sim4\%$. After de-redshifting all galaxy spaxels in the observed cube, it was verified that the recession velocity was consistent with zero, as expected for all the radial binned spectra. 
\subsection{Age and metallicity from spectral fitting}
Age and metallicity estimates, presented in Figure \ref{AgeZalpha}, obtained using the two different spectral fitting procedures STARLIGHT and pPXF, are found to be consistent with a central population age between 5 and 6 Gyr and a metallicity [Fe/H] of $-0.4$ dex. Even though small variations are seen between different methods, it should also be clear that quantities derived from mass- and luminosity-weighted methods should inherently differ due to different input information. Also, the fact that mass- and light-weighted properties, obtained from a single method, are slightly different means that the star formation history in NGC 1396 is extended (see e.g. \citealt{SanchezBlazquez2006a,SanchezBlazquez2006b}). The offset that is seen between STARLIGHT and pPXF computed luminosity-weighted ages is also due to the fact that STARLIGHT uses the information from the continuum in the fitting of a linear combination of SSPs and the reddening of the spectrum, while in the case of pPXF, the continuum is accounted for by a multiplicative polynomial.
With STARLIGHT, a negative extinction value $A_v$, was obtained for NGC 1396, which likely indicates that the flux calibration and/or the adopted extinction law is not ideal. This negative $A_v$ is compensated for by slightly older ages as seen in Figure \ref{AgeZalpha}.
\begin{figure*}
\begin{minipage}[c]{\textwidth}
\begin{center}
    \includegraphics[scale=0.4]{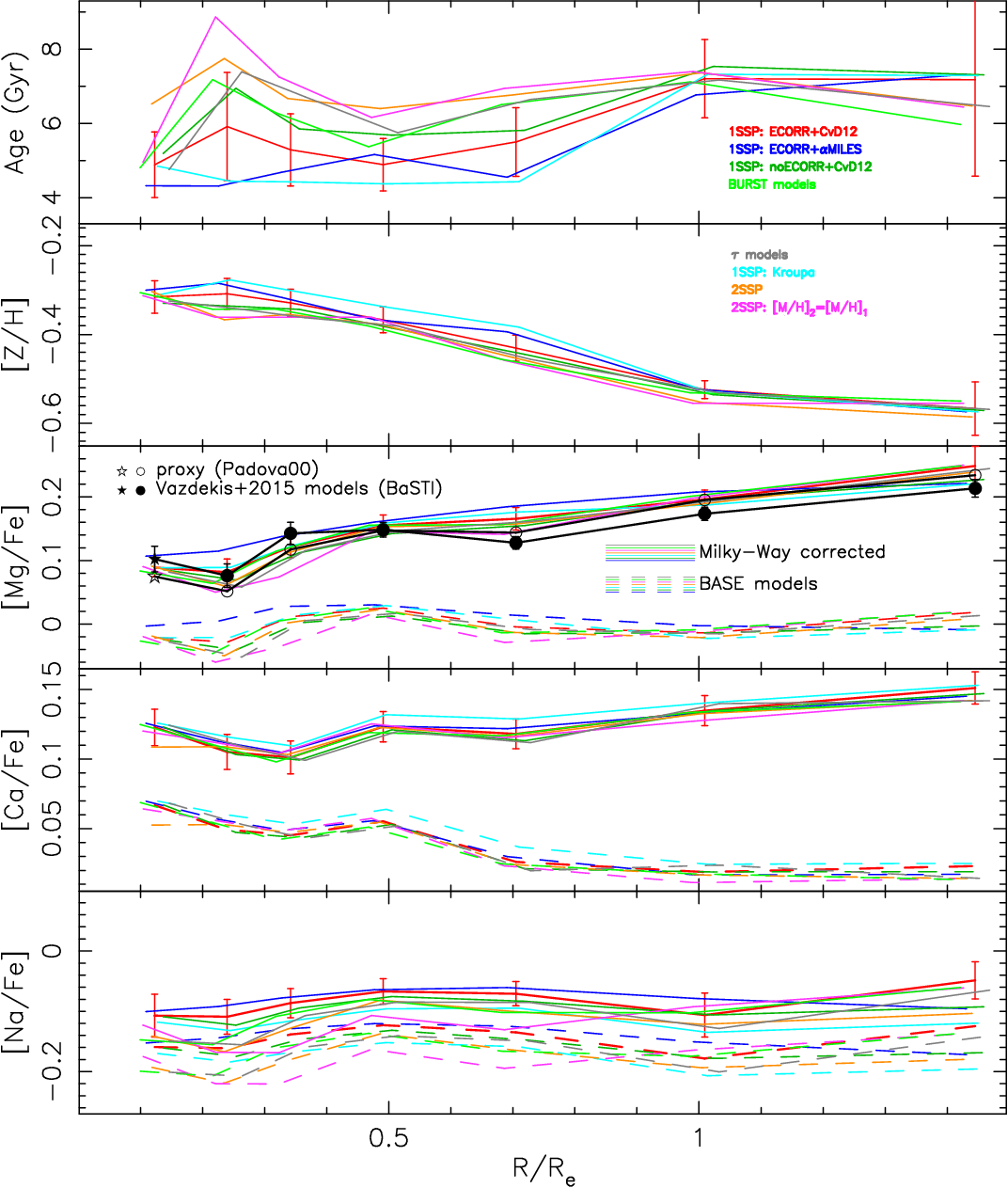}
    \caption{Best-fitting abundances as a function of galacto-centric radius obtained from index fitting. The parameters allowed to vary are age, [Z/H] and IMF slope, as well as [Ca/Fe], [Mg/Fe], [NaFe] from BASE-models. MW corrected abundances are indicated with solid lines while the abundances obtained from BASE-models are indicated by dashed lines. This first bin, contaminated by the NSC, is indicated by a star. The same colour coding applies to to this figure as explained for Figure \ref{radindex}.}
    \label{ab_gradients}
  \end{center}  
\end{minipage}
\end{figure*}
Apart from the statistical errors, the uncertainty is mainly dominated by the technique used to estimate the stellar population parameters and by the sky subtraction procedure. The uncertainty in the techniques was addressed by comparing two different methods, where differences of about 2 Gyr were seen. The relatively large scatter that we see in the age estimates, mainly comes from the (i) differences between light- and mass- weighted age estimates; (ii) degeneracy between IMF, SFH and age and, (iii) the different ways we treat the emission correction. This is a systematic error. When comparing results among the different radial bins, the errors to be adopted are much smaller and are given in Figure \ref{AgeZalpha}.

Because of the above arguments, in our opinion, there is no "more reliable" estimate, as different methods are well-known to have their specific drawbacks. Therefore we cannot prefer any method above another.

\subsection{Age and metallicity from index fitting}
Ages and metallicities obtained from the index fitting procedure, as shown in Figure \ref{ab_gradients}, display slightly different trends compared to full spectral fitting, which is in agreement with what might be expected when comparing these two methods. While the age shows a wide range for the radially binned spectra, which were modelled with different stellar populations as explained in Section \ref{sec:lineindices}, the metallicity shows a better constrained behaviour throughout all radial bins. Errors in age and metallicity estimates from the index fitting method are also found to be significantly larger than those found with the full spectral fitting method. This is due to the fact that absorption features and sky residuals affect the index measurements much more than fitting the full spectrum.       

\subsection{Elemental abundance ratios}  
In contrast to full spectral fitting, the index approach is focused on localised spectral regions to measure the effects of elemental abundances on spectral lines and line ratios. The index method is often less sensitive to absolute spectrophotometric calibrations and dust absorption, which makes it a good comparative method to complement full spectral fitting \citep{Stellarpopulations_IAUS262}. 
The MUSE wavelength range includes multiple emission lines and stellar absorption features useful for stellar population analysis. For this study, important features to consider are the gravity sensitive features: Mgb, NaD, NaI8190, TiO1, TiO2 and CaT. 

The radial behaviour of a number of measured indices is shown in Figure \ref{radindex} (black solid lines), which were fitted with models having a variety of star-formation histories, including one SSP, two SSPs and Burst and Tau models as explained in Section \ref{sec:lineindices}, on absorption line indices. The slight gradient observed in H$\beta_o$, implies that the central population is slightly younger, with an outward increase in age of $\sim(0.6 - 1.0)$ Gyr. This also agrees well with stellar population synthesis done with the STARLIGHT and pPXF codes for comparison (see top panel of Figure \ref{AgeZalpha}). Figure \ref{radindex} also shows model fits of one and two SSPs to the observed spectra, where indices from the models were compared to the radial behaviour of the measured indices. All indices are well fitted except for Ca1 in the outer regions of the galaxy. This is likely due to sky residuals found in the lower S/N Ca1 region of the outer radial bins. In general, no significant difference was found between one and two SSPs cases, thus a single SSP can be accepted as a good description of the selected spectral features.

\begin{figure*}
 \centering
 \includegraphics[width=21 cm ,height=18 cm,keepaspectratio=true]{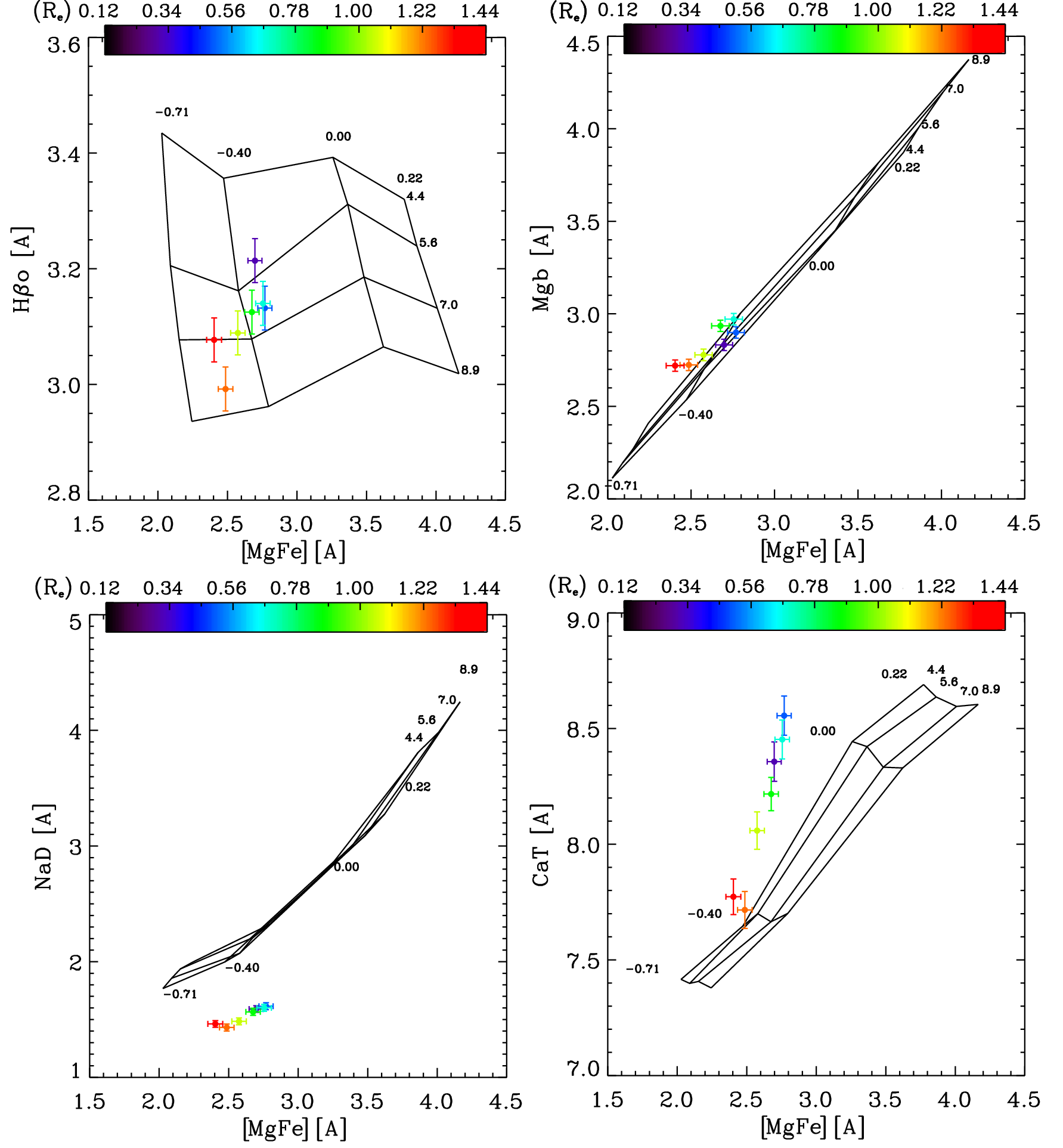}
 \caption{Index-index diagrams of H$\beta_o$, Mgb, NaD, and CaT as function of total metallicity [MgFe]. MIUSCAT models grids are plotted with IMF slope of 1.3, age ranging between 4.4 Gyr to 8.9 Gyr and metallicity ranging between $-0.71$ to $0.22$ dex. The colour coding indicates the radial distance in R/$R_e$ where red indicates the most distant radial bin from the centre.}
 \label{index1}
\end{figure*}

\begin{figure*}
 \centering
 \includegraphics[width=18 cm ,height=18 cm,keepaspectratio=true]{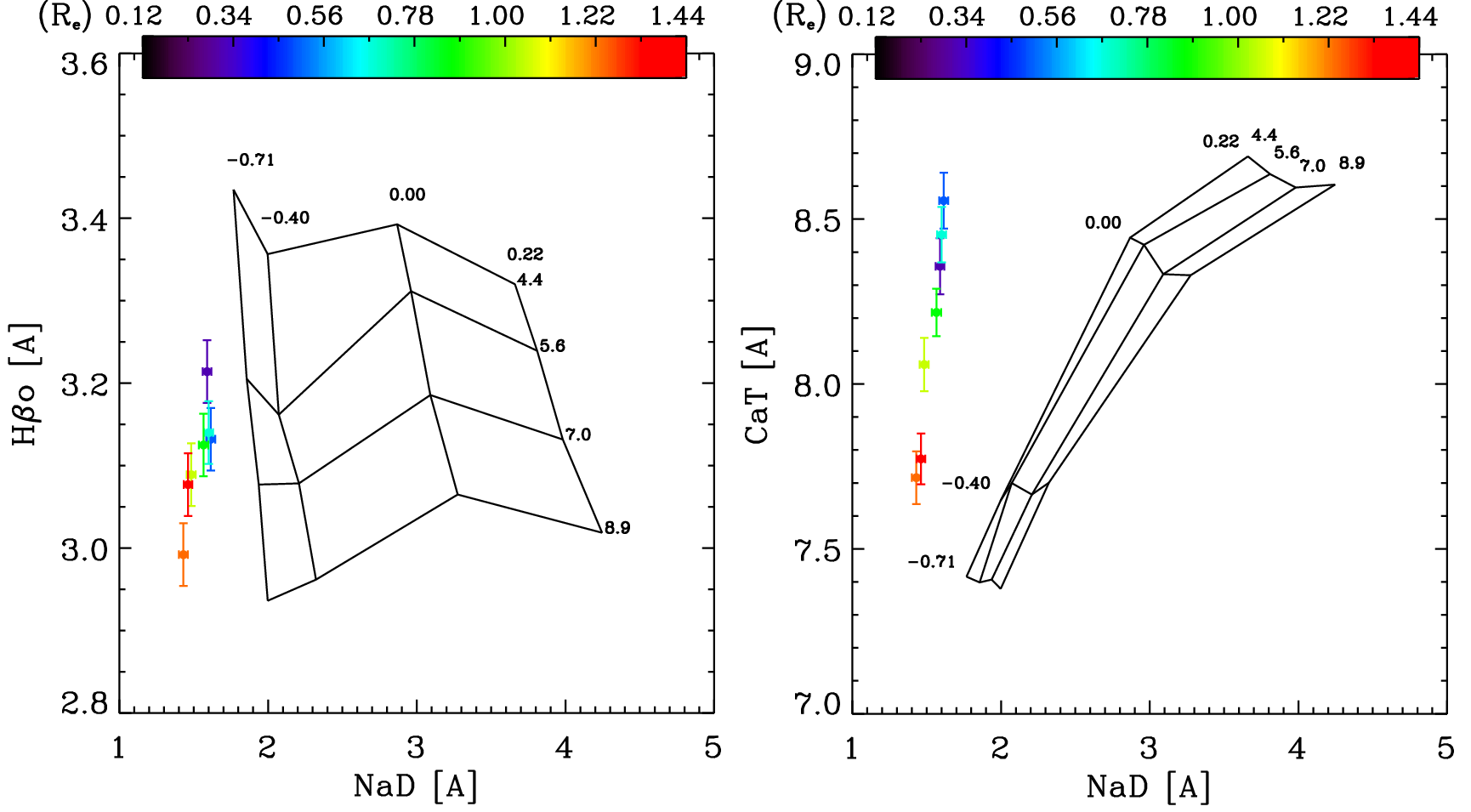}
 \caption{Index-index diagrams of H$\beta_o$ and CaT as function of NaD. MIUSCAT models grids are plotted with IMF slope of 1.3, age ranging between 4.4 Gyr to 8.9 Gyr and metallicity ranging between $-0.71$ to $0.22$ dex. The colour coding indicates the radial distance in R/R$_e$ where red indicates the farthest radial bin from the centre.}
 \label{index2}
\end{figure*}

Figure \ref{index1} shows index-index plots in which the gravity sensitive features are presented as a function of the total metallicity of the galaxy, [MgFe].
From the line strength measurements, highly elevated Ca values were found. CaT is shown in the lower right panel of Figure \ref{index1}, where a negative radial gradient is seen, with values decreasing from 8.5 \AA{} in the centre to 7.7 \AA{} in the outer regions. These values are strongly under-predicted by the over-plotted MIUSCAT models where a IMF slope of 1.3 was used together with metallicities between $-0.71$ and $0.22$ dex. It should be noted that the models have solar abundance ratios, which implies that the observations can only be explained by non-solar abundances.
Another interesting aspect observed in this galaxy is the very low Na abundance. NaD is shown to have a relatively flat profile with values between $\sim$1.6 \AA{} in the centre and $\sim$1.4 \AA{} in the outer region. NaD is also found to be much lower than the model grid in comparison to [MgFe], which could be explained by an under-abundance in Na compared to Fe. This can also be seen in Figure \ref{index2}, where the index-index diagrams is shown of H$\beta_o$ and CaT as function of NaD. The redshift of NGC 1396 is also large enough so that NaD is not affected by NaD in the Earth's atmosphere. NaI8190 shows comparable behaviour to that of NaD in the sense that it is also systematically below the solar-scaled grids at all radii.  
\begin{figure*}
\begin{minipage}[c]{\textwidth}
 \begin{center}
 \includegraphics[width=15 cm ,height=10 cm,keepaspectratio=true]{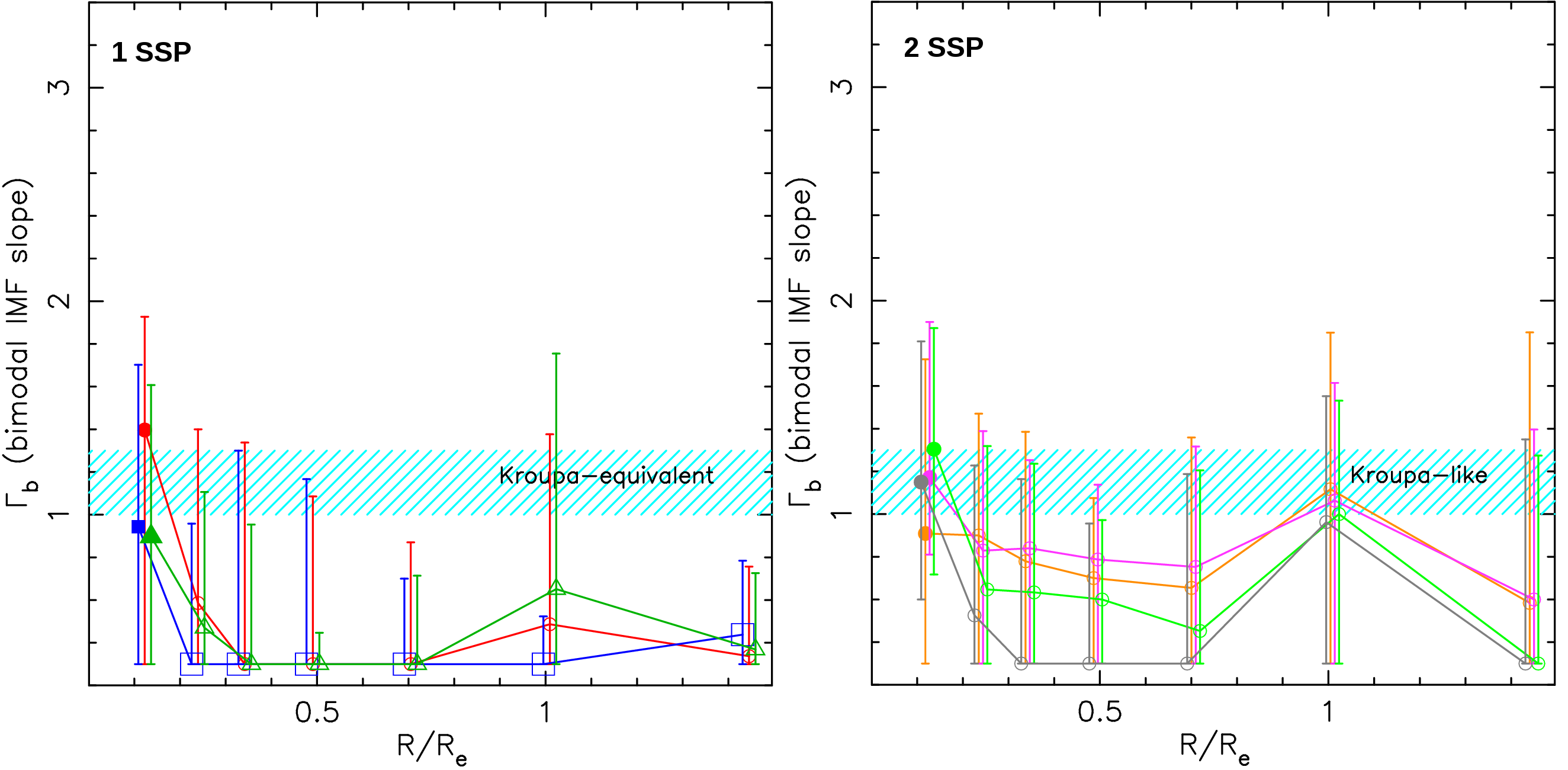}
 \caption{IMF slopes obtained for different realisations of a one SSP model (left) and two SSP models (right), as function of radial distance. Results are, within $1-2$ sigma, consistent with a Kroupa-like distribution, especially for the two SSP models, while the one SSP approach favours a more top-heavy distribution. Line colours are the same as in Figure \ref{radindex}}
 \label{IMF_slope}
 \end{center}  
\end{minipage}
\end{figure*}

In the case of the Burst and Tau decomposition, it was found that the solutions of both models populate similar regions as seen with the single SSP and two SSP models, implying that our results about radial trends of abundance ratios in NGC 1396 are robust against the adopted parametrisation of the SFH.
\subsection{Abundance gradients}
The behaviour of the colour and metallicity in NGC 1396 as a function of radius is as expected for a typical dE \citep{Koleva2009b, Koleva2011}. We see significant [Z/H] and [Mg/Fe] gradients of $-0.33 \pm 0.04$ and $0.20 \pm 0.02$ dex respectively, where the Ca and Na abundances seem to be much shallower. The gradients obtained from the simultaneous fitting of indices, together with age and metallicity gradients from spectral fitting, expressed as a function of $log$ r are tabulated in Table \ref{gradients}, and can also be seen for the full spectral fitting and the fits to the indices in Figures \ref{AgeZalpha} and \ref{ab_gradients} respectively. 

A younger population age is associated with the case where higher metallicities are produced during a more prolonged SFH. Therefore, the younger central population as deduced from the slight positive radial age gradient ($\sim$ 0.6-1.0 Gyr/$R_e$) is in agreement with the negative metallicity gradient. Metallicity is seen to decrease from the central region with a gradient of $\sim$0.21 dex per $R_e$, with a rather flat gradient inside $\sim0.4$ $R_e$. All other absorption line indices show negative gradients of which $<$Fe$>$ is the strongest at $-0.39 \pm 0.07$. NaI8190 appears to have a very flat gradient of $-0.06 \pm 0.02$ even though the gradient of NaD is shown to be more significant at ($-0.18 \pm 0.04$). This is consistent with the fact that NaD is significantly more sensitive to total metallicity than NaI8190, and the fact that NGC 1396 exhibits a negative metallicity gradient \citep{Spiniello2012}. 
\begin{table}
\begin{center}
\caption{Gradients of abundances and line indices. For comparison, values with (*) indicate the gradients obtained by using the method of spectral fitting from STARLIGHT, pPXF mass- and luminosity weighted}
\begin{tabular}{lcc}
\hline
\hline
\bf{Parameter} & \bf{Gradient} $(\Delta(\mathrm{par})/\Delta(\log(r))$ & \\
Age [Gyr]&$2.011 \pm 0.745$ & \\ 
Age *(STARLIGHT) [Gyr]&$1.571 \pm 0.268$& \\ 
Age *(pPXF M-weighted) [Gyr]&$3.308 \pm 0.496$& \\
Age *(pPXF L-weighted) [Gyr]&$2.999 \pm 0.354$& \\
Colour [mag (F475-F814)]&$-0.077 \pm 0.006$& \\
\hline
\textit{\bf{Abundances:}}& & \\
Z/H&$-0.334 \pm 0.044$& \\
Z/H *(STARLIGHT)&$-0.455 \pm 0.019$& \\
Z/H *(pPXF M-weighted)&$-0.432 \pm 0.031$& \\
Z/H *(pPXF L-weighted)&$-0.444 \pm 0.020$& \\
Ca/Fe&$0.059 \pm 0.012$& \\
Mg/Fe&$0.201 \pm 0.016$& \\
Na/Fe&$0.051 \pm 0.028$& \\
\hline
\textit{\bf{Indices:}}& & \\
Hb$_o$&$-0.125 \pm 0.061$& \\
Mgb&$-0.324 \pm 0.106$& \\
$<$Fe$>$&$-0.544 \pm 0.019$& \\
NaD&$-0.252 \pm 0.041$& \\
NaI8190&$-0.063 \pm 0.025$& \\
Ca1&$-0.211 \pm 0.039$& \\
Ca2&$-0.351 \pm 0.040$& \\
Ca3&$-0.243 \pm 0.046$& \\
\hline
\end{tabular}
\label{gradients}
\end{center}
\end{table}
\subsection{IMF}
In Figure \ref{IMF_slope}, the best-fit IMF slopes are shown as a function of galacto-centric radius. By using a bimodal IMF, which is characterised by its slope, $\Gamma_b$, as a single free parameter (see \citealt{LaBarbera2013}), a slope of $\Gamma_b= \sim1.3$, could be seen as a good representation of a Kroupa-like distribution, while a top-heavy and bottom-heavy IMF is consistent with $\Gamma_b <1$ and $\Gamma_b > 1.3$ respectively. In Figure \ref{IMF_slope}, the Kroupa-equivalent region is indicated, which is the region where the bimodal IMF is essentially equivalent, in terms of M/L and mass fraction in low-mass stars, to a Kroupa distribution.

By comparing these fitted IMF slopes, in each radial bin, it is seen from the one SSP and two SSP models that the two SSP slopes (right panel of Figure \ref{IMF_slope}) are larger than the single SSP cases (left panel of Figure \ref{IMF_slope}), with a median value of $\Gamma_b\sim0.8$. All bins fitted with the two SSPs are consistent within 1 $\sigma$ with a Kroupa-like IMF \citep{Kroupa2001}. This could be explained by the fact that most IMF-sensitive features have similar behaviour with both age and IMF slope, which means that with a younger SSP contribution (as in the two SSP case), one tends to infer a larger IMF slope. However, in the case of giant elliptical galaxies, where a lower contribution of a young population is expected, two SSP models give very consistent results regarding the IMF slopes with respect to one SSP models (see \citealt{LaBarbera2013}).

\section{Discussion and conclusions}
This is the first time that Integral Field Spectroscopy is published for a dwarf elliptical in the near infrared region. Previous work on dwarfs \citep{Hilker2007, Koleva2009a, Koleva2009b, Koleva2011, Toloba2014}, used Echelle and long-slit spectroscopy, or observed bluer spectral regions, using SAURON IFU data \citep{Rys2013, Rys2015}.

In this discussion there will be concentrated on our new results in the area of abundance ratios and star formation histories, the IMF and stellar population gradients and their implications on how dwarf ellipticals form and evolve.
\subsection{Elemental abundance ratios}
One of the most significant results obtained from this study is the unusual abundance ratios observed in Ca and Na when compared to massive elliptical galaxies and the Local Group. Abundance ratios in a specific environment can be used to gain more information on the IMF, SFH as well as time scales for chemically evolving systems \citep{McWilliam1997}. Until now, abundance ratios in the Milky Way, GCs and galaxies in the Local Group have been documented extensively e.g. \citep{Tolstoy2007, Kaufer2004, Piotto2009} in order to understand small and large scale environmental variations.
Nowadays we are also able to measure abundance ratios of many elements from integrated light. This is done by fitting stellar population models, using empirical stars, modified using theoretical responses of abundance ratio variations to spectra \citep{Walcher2015,Coelho2014}. \citet{Conroy2014} used this method to derive abundance ratios for high S/N spectra of early-type galaxies as a function of mass.
Abundance ratios can provide important clues for the mass assembly of galaxies by using GCs. It has been shown that GCs in our galaxy, and recently also in M31, contain a significant fraction of stars with different abundance ratios of light elements (Li, C, N, O, Na, Mg, and Al) compared to field stars \citep{Gratton2004, Beasley2005}. For example anti-correlations between Na and O and between Mg and Al have been found, where it is believed that Na is made out of Ne and Al from Mg. Since this is occurring only in GC stars, the environment in which the stars were formed, must have been important. Here we investigate whether similar effects are seen in a dwarf galaxy, and whether this helps us in understanding this phenomenon. 

Figure \ref{ab_comp} shows a quantitative comparison between the Ca, Mg and Na abundances found in NGC 1396 and those observed in our local environment. By making a comparison between the local (resolved) and more distant (unresolved) systems, studied with integrated light, it is important to understand how accurate SFHs can be recovered and compared to those obtained from the most accurate method of constructing a colour-magnitude diagram. In a study by \citet{RuizLara2015}, they show that a very good agreement is indeed found in recovering the SFH and age-metallicity relation from the two methods i.e. colour magnitude diagrams and integrated spectral analysis. 
Compared to abundances found in Local Group dwarfs and LMC stars, it is seen that the light elements Na and Mg in NGC 1396 are both more overabundant, while Na is under-abundant in comparison with MW stars. However, the Ca, relative to Fe, abundance is closer to local observed abundances for lower metallicity environments ($-2.0 <$ [Fe/H] $< 0.0$), which is found in the LMC GCs by \citet{Colucci2012}, with a mean value of ($0.29 \pm 0.09$) dex.
\subsubsection{Ca and Mg abundances}
The CaT lines constitute a very interesting albeit not well studied spectral region also containing the underlying H Paschen absorption (PaT). CaT is a good metallicity indicator in lower than solar metallicity environments, but in giant ellipticals CaT is much weaker than predicted by models \citep{Cenarro2003,Cole2005,Sakari2016}. \citet{Cenarro2003} suggested that the IMF in such galaxies was different from the solar neighbourhood, more bottom-heavy, which is one of the factors which has led to a discussion about the IMF depending on mass or velocity dispersion in galaxies (see \citealt{Conroy2013}). However in dEs, given their low metallicities, CaT should be in principle a good metallicity-indicator.
\begin{figure*}
\begin{minipage}[c]{\textwidth}
\begin{center}
    \includegraphics[scale=0.6]{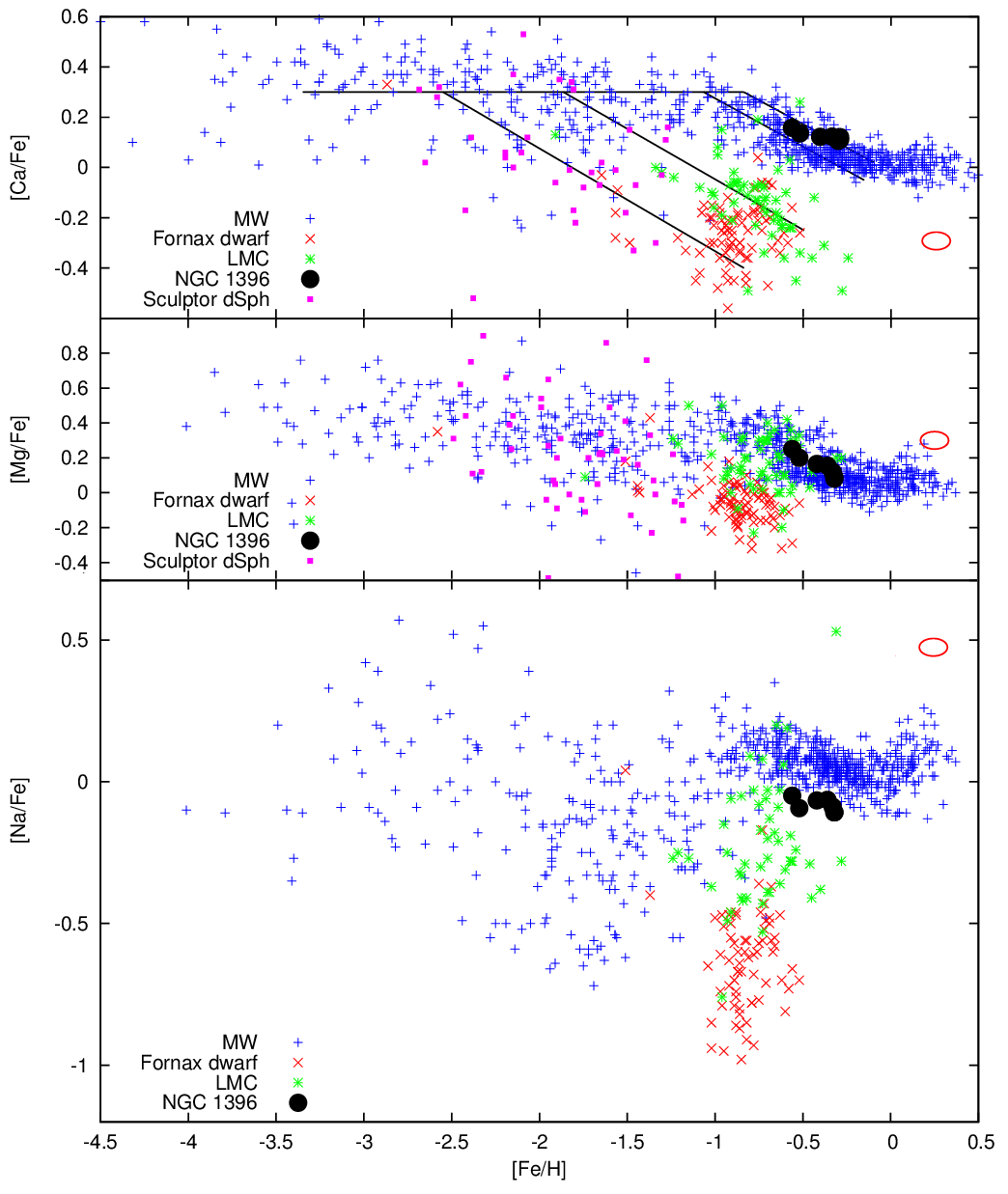}
    \caption{Comparison of Ca, Mg and Na abundances (corrected for the MW abundance pattern, see Section \ref{ab_correction}) as a function of metallicity, [Fe/H], between the Fornax local dwarf (shown in red) \citep{Tolstoy2009}, the LMC red giants (shown in green) \citep{Pompeia2008}, the Milky Way, indicated in blue \citep{Venn2004a}, the Sculptor dSph, in pink \citep{Kirby2009}, and NGC 1396 (black circles). The red ellipses in both panels indicate the region of massive elliptical galaxies \citep{Henry1999,Worthey2011,CvD12}. Errors for NGC 1396 are of the same order as the symbol. The solid lines in the top panel indicate a visual trend observed in Ca abundance as function of [Fe/H]}
    \label{ab_comp}
  \end{center}  
\end{minipage}
\end{figure*}

Recently \citet{Zieleniewski2015} found evidence of a young stellar component in M32 using far red stellar absorption features. Even though M32 is a compact elliptical, while NGC 1396 is not, it is still interesting to do a comparison between two early-type galaxies with very similar masses of about $10^9$ M{$_\odot$}. It was found that the CaT index measured in the central 10 pc of M32 ($\sim$8 \AA{}) is very high in comparison with M31 ($\sim$7.5 \AA{}), which was covered by multiple pointings in the same study. Na measurements from the same study also show a behaviour very different from M31, where [Na/Fe] in M32 was found to be significantly lower than in M31.

For NGC 1396, the measured CaT values are considerably higher than the model prediction (Figure \ref{index1}). This indicates that Ca is overabundant with respect to Fe, rather than under-abundant, as seems to be the case in giant ellipticals \citep{Vazdekis1997,Saglia2002,Cenarro2003}. The reason why the Ca-abundance is so high is not very clear yet. Ca is seen to be slightly overabundant in the [Ca/Fe] vs [Fe/H] plot of Figure \ref{ab_comp}, where the [Ca/Fe] measurements of NGC 1396 appear to be between those of the MW disc stars ([Ca/Fe=0]) and the older thick disc stars ([Ca/Fe] $>0$), although much closer to the disc stars.  

Due to the strong sensitivity of CaT to changes in [Ca/Fe] and [Mg/Fe], and the observed weakening of the optical and near infrared Na indices with increasing [Mg/Fe] in giant ellipticals \citep{CvD12}, an under-abundant [Na/Fe] or [Mg/Fe] would thus also imply higher CaT values. This observed anti-correlation of CaT lines with [Na/Fe] and [Mg/Fe] together with a positive [Ca/Fe], is sufficient to explain all Ca line-strengths for NGC 1396.  

The idea was also tested whether a strong CaT index, as previously also noted by \citet{Kormendy1999} and \citet{Cenarro2008}, could be explained by the inclusion of a super-young stellar component ($<100$ Myr), but it was found that the amount of young populations needed would be so large that e.g. the Balmer lines could not be fitted. Together with this, we also analysed the separate effect on the three CaT lines (Ca1+Ca2+Ca3). It was found that all three lines increase with decreasing age, for ages $< 10$ Myr, while Ca2 increases the least and Ca1 the most of the three lines. We also see, for NGC 1396, that Ca2 has the largest discrepancy from the single SSP index-index grid, while Ca1 is almost on top of the grid. This is exactly the opposite than what would be expected if there were a super-young component driving CaT up. Because of this, when trying to fit 2SSP models, by including super-young ages, we never found a significant contribution from the super-young component, and the estimate of [Ca/Fe] is very similar to what we show in this paper. 

Another possibility is that the overabundant Ca could point to a more top-heavy IMF distribution in the dwarfs, which would be expected if Ca is mostly produced in more massive k- and M-type stars. However [Ca/Fe] increases outwards for NGC 1396, in a similar way as [Mg/Fe], which means that the Ca abundance is also connected to the duration of star-formation (as [Mg/Fe]) in this dwarf, rather than a top-heavy IMF.

It is expected that Ca follows the abundances of other $\alpha$-elements due to their common nucleosynthetic origin \citep{Wheeler1989,Worthey2011}. However, with later observations it became clear that not all $\alpha$-elements are enhanced in lockstep and that Ca and Mg abundances are not correlated for giant ellipticals \citep{Vazdekis1997,Thomas2003b}. In NGC 1396, both Ca and Mg are enhanced in the centre, and show a similar radial trend, although there is some evidence that Ca might be higher compared to Mg in the outermost regions. This might be explained by the distinction which is seen in the $\alpha$-elements between O and Mg on the one hand and Si, Ca and Ti on the other. O and Mg are known to be produced during the hydrostatic He burning phase in massive stars before the onset of the SNII explosion, during which their yields are not affected by SNII explosions, however, Si, Ca and Ti, are mostly produced during the SNII explosion \citep{Henry1999,Tolstoy2009}.

The enrichment history of a stellar system is mostly determined by the ratio and timescale of the initial SNII enrichment, which is followed by SNIa \citep{Matteucci2001} and it is known that the lifetime of SNIa progenitors are longer than that of SNII. In a dwarf-type system that cannot retain the metals, mostly produced by SNII, and lost by feedback mechanisms, [$\alpha/Fe$] will decrease with the onset of SNIa \citep{Schiavon2010,Hill2012}. Probable scenarios which could explain this phenomenon have been proposed which include short star formation timescales, top-heavy IMF and selective winds, where all of these processes depend on the fact that Mg is produced by Type II supernova while most Fe is produced in Type Ia supernova \citep{Faber1992}. In NGC 1396, [Mg/Fe] and [Ca/Fe] is seen to be enhanced, while both increase outward. This indicates that, relative to the SFH, the onset of SNIa started at a similar time scale than compared to what is observed in the Milky Way. This indicates that the build-up of stellar populations in NGC 1396 has been rather disc-like, supporting the evolutionary trend in dwarf irregular galaxies of a disc-like formation in comparison with the enrichment history of the MW disc. 

\begin{figure*}
\begin{minipage}[c]{\textwidth}
\begin{center}
    \includegraphics[scale=0.45]{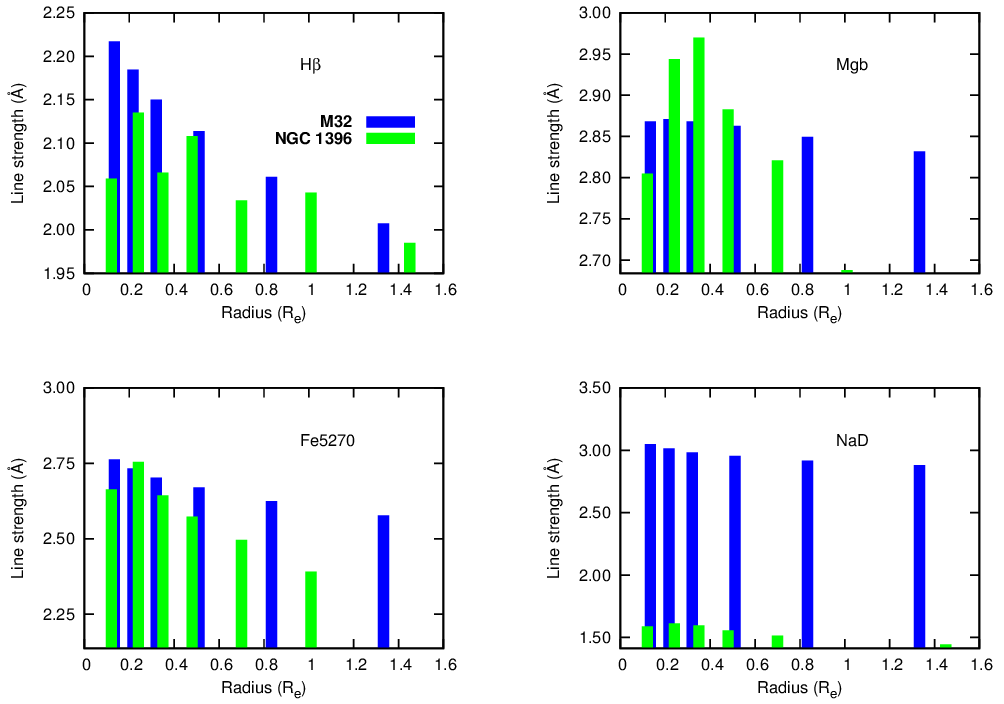}
    \caption{Radial comparison of the Lick indices H$\beta$, Mgb, Fe5270 and NaD, between NGC 1396 and M32. Line strengths from M32, measured at Lick resolution, are obtained from \citet{Worthey2004}, and converted to LIS 5.0 \AA{}.}
    \label{in_comp}
  \end{center}  
\end{minipage}
\end{figure*}

In a comparison between M32 and NGC 1396, shown in Table \ref{indices} and in Figure \ref{in_comp}, it can be deduced from the age sensitive H$\beta$ and Fe indices that M32 is younger and more metal rich respectively, whereas from individual line strength measurements, these two systems show similar elevated CaT and low NaI8190 line strengths. In a comparison between the Mg and Fe indices for M32 and NGC 1396, it is seen that the Fe and Mg indices are consistently stronger in M32. In Table \ref{indices} and Figure \ref{in_comp}, we compare line strengths of spectral indices, measured from the radially binned spectra of NGC 1396, with those measured at similar galacto-centric distance for M32 (taken from \citet{Worthey2004}). In the Figure, values for NGC 1396 are plotted in green and those from M32 in blue. The indices from \citet{Worthey2004}, were measured at Lick resolution, which was converted to the LIS 5.0 \AA{} system by using the conversion factors as presented in \citet{Vazdekis2010}.
\begin{table*}
\tiny
\centering
\caption{Comparison of measured indices as function of radial galactic distance between NGC 1396, from this study, and M32 from \citet{Worthey2004}}
\label{indices}
\begin{tabular}{lllllllllllll}
\hline
\hline
\multicolumn{12}{c}{Lick indices measured in NGC 1396 and M32} \\
R$_e$ & H$_\beta$ & Fe5015 & Mg1 & Mg2 & Mg$_b$ & Fe5270 & Fe5335 & Fe5406 & Fe5709 & Fe5782 & NaD \\
\hline
\textbf{NGC 1396} & & & & & & & & & & & \\
0.12  &  2.059 &  4.442 &  0.032 & 0.139 &  2.805 & 2.664 &  2.233 & 1.506 & 0.794 & 0.638 & 1.590 \\
0.24  &  2.135 &  4.352 &  0.033 & 0.143 &  2.944 & 2.755 &  2.225 & 1.545 & 0.812 & 0.662 & 1.613 \\
0.34  &  2.066 &  4.257 &  0.033 & 0.142 &  2.970 & 2.644 &  2.273 & 1.566 & 0.771 & 0.665 & 1.597 \\
0.49  &  2.108 &  4.149 &  0.031 & 0.137 &  2.883 & 2.573 &  2.163 & 1.451 & 0.741 & 0.628 & 1.557 \\
0.70  &  2.034 &  3.914 &  0.031 & 0.133 &  2.821 & 2.496 &  2.067 & 1.411 & 0.685 & 0.549 & 1.516 \\
1.01  &  2.043 &  3.912 &  0.028 & 0.126 &  2.688 & 2.391 &  2.043 & 1.326 & 0.717 & 0.528 & 1.412 \\
1.45  &  1.985 &  3.709 &  0.031 & 0.127 &  2.684 & 2.136 &  2.006 & 1.332 & 0.666 & 0.442 & 1.445 \\
\hline
\hline
\textbf{M32} & & & & & & & & & & & \\
0.007  & 2.289 &  4.830 &  0.068 & 0.183 &  2.775 & 2.834 &  2.547 & 1.739 & 0.924 & 0.829 & 3.177 \\
0.026  & 2.285 &  4.736 &  0.067 & 0.179 &  2.821 & 2.826 &  2.534 & 1.740 & 0.936 & 0.825 & 3.156 \\
0.052  & 2.264 &  4.689 &  0.066 & 0.177 &  2.849 & 2.806 &  2.510 & 1.728 & 0.933 & 0.816 & 3.113 \\
0.078  & 2.247 &  4.660 &  0.066 & 0.175 &  2.860 & 2.790 &  2.491 & 1.718 & 0.928 & 0.809 & 3.085 \\
0.133  & 2.217 &  4.623 &  0.065 & 0.174 &  2.868 & 2.763 &  2.460 & 1.699 & 0.919 & 0.796 & 3.049 \\
0.217  & 2.185 &  4.590 &  0.064 & 0.173 &  2.871 & 2.733 &  2.427 & 1.677 & 0.908 & 0.781 & 3.016 \\
0.333  & 2.150 &  4.560 &  0.064 & 0.172 &  2.868 & 2.703 &  2.392 & 1.655 & 0.896 & 0.767 & 2.984 \\
0.500  & 2.114 &  4.533 &  0.063 & 0.170 &  2.863 & 2.670 &  2.356 & 1.631 & 0.882 & 0.752 & 2.956 \\
0.833  & 2.061 &  4.497 &  0.062 & 0.169 &  2.850 & 2.624 &  2.304 & 1.597 & 0.861 & 0.730 & 2.917 \\
1.333  & 2.008 &  4.464 &  0.062 & 0.167 &  2.832 & 2.578 &  2.252 & 1.562 & 0.839 & 0.708 & 2.882 \\
\hline
\end{tabular}
\end{table*}

\subsubsection{Na abundances}
In comparing our NaD and NaI8190 indices to those found by \citet{Zieleniewski2015}, our values of NaD and NaI8190 are both found to be much lower than those of the MIUSCAT model grids. In the case of the more compact dwarf system M32, NaD from \citet{Zieleniewski2015} are consistent with the solar-scaled grid, while NaI8190 is too low compared to the MIUSCAT model grid. However, the NaD measurements for M32 \citep{Zieleniewski2015} is still significantly higher that what is found for NGC 1396. 

[Na/Fe] and [Ca/Fe] are usually found to have an opposite behaviour compared to [Fe/H] (see \citealt{Tolstoy2009,Worthey2011} and references therein). This anti-correlation is believed to be caused by the contribution of SNII, which are mostly responsible for the production of $\alpha$-elements, which in turn strongly influence the pseudocontinua of especially the NaI8190 index, and also by the metal dependent yields \citep{Tolstoy2009,CvD12}. An interesting question that arises when considering the Na abundance in this galaxy is how [Na/Fe] can be as low as $\sim-0.1$ dex when the $\alpha$-element abundance is found to be larger than zero. Together with the fact that NaI8190 is over-predicted by the MIUSCAT models in M32 \citep{Zieleniewski2015}, it has been found that Na is also significantly de-enhanced in the Fornax dSph \citep{Letarte2010}. However, the reason for this de-enhancement in Na, found mostly in dwarfs, is still not clear. \citep{Conroy2014} suggest that, for giant early-type galaxies, [Na/Fe] correlates with galaxy mass/velocity dispersion. By adding NGC 1396, the LMC and the Fornax dwarf, we show that the correlation covers a much larger range in mass and velocity dispersion. This trend can also be seen from Figure \ref{ab_comp}.

According to \citet{Kobayashi2006}, [Na/Fe] and [Al/Fe] in metal-poor MW stars are lower compared to solar abundances by $\sim1.0$ and $0.7$ dex respectively. It seems that SN yields for odd-Z elements, such as Al and Na, are the only elements that have a significant dependence on metallicity \citep{Kobayashi2006}. From the slow chemical enrichment by SNIa, significantly less Na compared to Mg might be expected in low metallicity dwarfs. The contrary also applies and might therefore explain why strong Na enhancement is seen in metal-rich ETGs \citep{Smith2015}(R\"{o}ck et al. 2016 in prep).

\subsection{Star formation histories}
In the past it has also been shown that dEs have a large range of ages (between $\sim3$ and $\sim12$ Gyr), obtained from measurements of age sensitive indices such as H$\beta$ \citep{Michielsen2008r, Koleva2009a, Rys2015}.
Intermediate ages with a shallow age gradient is seen from the central population outwards. This together with photometry results on NGC 1396, points to the presence of a younger central population or nuclear star cluster. Notice that in earlier work, a certain fraction of dwarf galaxies in all environments in the nearby universe were found to have blue core regions \citep{Gu2006,Tully2008}. Nuclear star clusters, with younger populations, have also been found in a large fraction of the dwarf galaxies \citep{DenBrok2011,Turner2012,Georgiev2014}. This has also been seen from UV-optical colours of dwarfs in the Virgo cluster \citep{Kim2010}. Although blue cores have been found in dwarfs, the question still remains whether they are formed during the last stages of star formation or whether it could be explained by re-accreted gas as proposed for the galaxy (FCC 46) presented in \citet{FCC046}. Unfortunately, even with the excellent spatial resolution provided by MUSE, it is not sufficient, in the case of NGC 1396, to constrain the stellar population properties of the NSC.  

It is also important to note that consistent abundance ratios for NGC 1396 are found by using one SSP and two SSPs, as well as models with extended (e.g. exponentially declining) SFHs. By only using one SSP, a degeneracy might be expected between the abundance ratios and the galaxy star-formation history. Indeed, our results indicate no significant degeneracy and therefore a robust estimation of the abundances. 

From the spectral fitting we found evidence that the star formation activity has been going on for an extended period of time. This could be deduced from the fact that a significant difference was noticed between the luminosity- and mass weighted ages from pPXF (see Figure \ref{AgeZalpha}).  
\subsubsection{IMF} 
Analysis of gravity sensitive features points to a dwarf-to-giant ratio in the IMF which is consistent with either a top-heavy or a Kroupa-like distribution. This is in stark contrast to the more bottom-heavy distributions preferred in populations of massive early-type galaxies, where the contribution of low-mass stars are much more important \citep{Conroy2012b,Spiniello2012,LaBarbera2013,Spiniello2014,LaBarbera2016}. 
By fitting two populations, it is found that the results are consistent with the conventional Kroupa IMF, which also indicates that the SFH can affect the measurements of the IMF. This however, is only true for the case of a top-heavy or Kroupa-like distribution, because for a bottom-heavy distribution, the dependence of gravity-sensitive features on the IMF slope is much stronger, and overwhelms that of the adopted SFH as also shown by \citet{LaBarbera2013}.
From our fits it was concluded that it is not needed to discard the classical Kroupa IMF, however, a bottom-heavy IMF for this system can be excluded at all radii, even allowing for a wide range of SFHs.

The only study which has addressed the normalisation of the IMF in dEs \citep{Tortora2016}, using an hybrid approach based on the mass discrepancy between the dynamically inferred mass and the Chabrier stellar mass, shows that dEs might have a large scatter in their mass-to-light normalisation which range from a bottom-light to a (either top-or bottom-) heavy IMF. However, for NGC 1396 it is found that the IMF is consistent with a Kroupa or top-heavy distribution. We expect to investigate the broader correlation with galaxy properties (see e.g. \citet{Tortora2016}, on a larger sample using gravity sensitive indices with forthcoming MUSE data.

\subsection{Gradients}
Gradients in dwarfs are presumably produced as a result of outside-in formation \citep{DenBrok2011}, where the star formation is initially spread out through the galaxy and evolves in a burst-like mode, contracting to the centre of the galaxy. This process has also been described using hydrodynamical simulations of isolated dwarf galaxies, in which \citet{Valcke2008} showed, by following the evolution of a collapsing gas cloud in an existing dark matter halo, that successive starburst episodes could restrain the star formation to a smaller central region. 
This process could then be responsible for a positive age gradient to appear in the galaxy together with positive colour and abundance gradients. The fact that we observe a shallow, although still positive, abundance gradients means that we are most probably dealing with a single component, in comparison to the MW thin disk, thick disk and bulge.

In the past claims were made that positive age and colour gradients had been found in dEs, but according to a study of \citet{DenBrok2011} in the more massive Coma cluster, this is most likely a result of the inclusion of the nuclear regions in the colour profiles, where one expects to find the younger stars, in e.g. the nuclear star clusters in dwarfs as is also present in the blue core of NGC 1396.
From the literature it is clear that not many dwarfs have indeed been found with a positive colour gradient (see e.g. \citealt{Spolaor2009}), from which \citet{DenBrok2011} also deduce that the influence of metallicity on the colour gradient of the galaxy is much more profound than that of the age gradient and could in principle be disentangled by combining optical and NIR colour gradients \citep{FalconB2011,LaBarbera2011}. 

The measured metallicity and colour gradients of NGC 1396 compares well to those of \citet{DenBrok2011}, for galaxies in the Coma cluster. The metallicity gradient of $-0.33$, for NCG 1396, is in agreement with the average metallicity of $-0.30$ for the 21 Coma dwarfs in the sample by \citet{SanchezBlazquez2006b}. In other comparisons with massive ellipticals from \cite{Raskutti2014} and \cite{Greene2015} and simulations from \citet{Kobayashi2004} and \citet{Hirschmann2015}, we find a similar range in the measured metallicity gradients, measured from $0.5$ $R_e$ to $1.5$ $R_e$ for comparison, (\citealt{Raskutti2014}:$-0.372 \pm 0.032$; \citealt{Greene2015}:$-0.472 \pm 0.074$; NGC 1396: $-0.445 \pm 0.048$, whereas the age gradients in the massive ellipticals are mostly negative and slightly steeper. We therefore find no apparent trend with galaxy mass for this particular dwarf galaxy. This is also in agreement with the findings from \citet{Kobayashi2004}, who also found no correlation between metallicity gradients and galaxy masses from simulations. 

Metallicity gradients similar to that of NGC 1396 have also been observed by \citet{Koleva2009a} who studied 16 dEs of which 10 showed strong metallicity gradients between $-0.3$ and $\sim-0.6$. It should be noted that those gradients are strongly affected by the age-metallicity degeneracy, which is why it is also important to compare gradients in individual indices.

\begin{figure}
\begin{minipage}[l]{8.5 cm}
\begin{center}
 \includegraphics[width=10 cm ,height=5 cm,keepaspectratio=true]{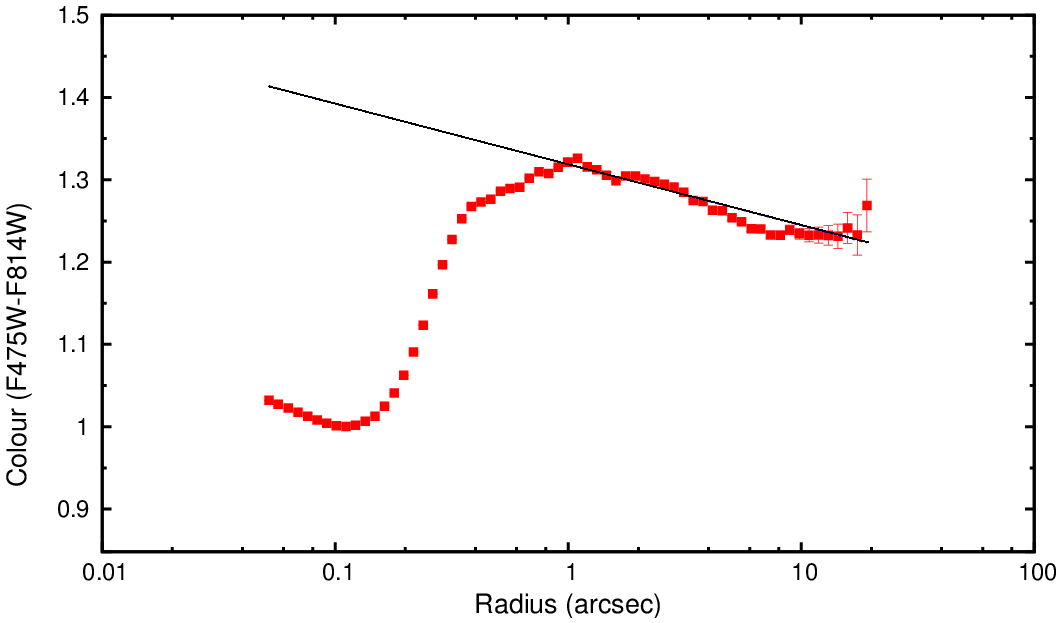}
 \caption{Colour Profile of NCG 1396 from HST archive data, with a gradient fit of $-0.077 \pm 0.006$ to the region outside of the central 1 arcsec. The central 0.2 arcsec is dominated by a nuclear star cluster.}
 \label{ColourProfile}
 \end{center}  
\end{minipage}
\end{figure}

From the colour profile of NGC 1396 (see Figure \ref{ColourProfile}) a blue central core region is visible, which is also in agreement with what is found from the photometric analysis (Hamraz et al (2016). in prep). The colour gradient (excluding the nuclear cluster region) measured from NGC 1396 ($\sim-0.08$) [F475-F814] agrees well with the average colour gradient measured for the sample of galaxies in the Coma cluster, with the same magnitude range, which is found to be $\sim-0.06$.
From a simulations point of view, \citet{Schroyen2013} concluded that metallicity gradients in dwarf galaxies are gradually built up in non-rotating dwarfs and can survive for extended periods of time, while only being affected or weakened by external disturbances. This is also in agreement with a possible scenario where the progenitor of NGC 1396 could have been a slow rotating dwarf irregular system, that entered the Fornax cluster. For a range in mass from $10^7$ to $10^9$ M{$_\odot$}, the range of metallicity gradients predicted by simulations by \citet{Schroyen2013}, ($-0.57$ to $-0.01$ dex/kpc) also correspond to the value obtained for NGC 1396 measured as $-0.394 \pm 0.041$ dex/kpc (measured within 1kpc outside the central two radial binned region for direct comparison). 

\section{Summary}
\begin{itemize}
 \item We presented stellar populations and abundance ratios obtained from MUSE IFU data of the dE NGC 1396 in the Fornax cluster. From deep, high spatial resolution, IFU spectroscopic data we are able to study the chemical abundances, the star formation history and the initial mass function as a function of galacto-centric distance. 

 \item By studying red absorption features for the first time in a typical dE, we find unusually overabundant values of [Ca/ Fe] $\sim$ +0.1, and under-abundant Sodium with [Na/ Fe] values around $-0.1$. Together with Ca, Mg is also enhanced and both are showing a positive radial gradient. We compared the Na abundances from this system to that found in massive elliptical galaxies and also to dwarf galaxies in the Local Group, from which we find a correlation between [Na/Fe] and [Fe/H], going all the way from the Fornax dwarf galaxy to giant elliptical galaxies. The current best explanation is that Na enrichment yields are strongly metal dependent. 
 
 \item We found evidence that the star formation activity has been going on for an extended period of time, where the abundance estimates are consistent with a disc-like build-up of the stellar populations of NGC 1396. This is also evident from the fact that a significant difference is seen between the luminosity- and mass weighted ages from pPXF.   

 \item We have fitted our radially binned spectra with a variety of stellar population models, in order to constrain the IMF. With this we are able to firmly rule out a bottom-heavy distribution for this system where the IMF is consistent with either a Kroupa-like or a top-heavy distribution.
 \end{itemize}

\section{Acknowledgements} 
The research was enabled in part by support provided by the National Research Foundation of South Africa and the EU-SATURN program to JM. We thank the reviewer for his/her careful reading of the manuscript and helpful comments.
The observation of this publication are based on a proposal from the FDS team. J.\textasciitilde F-B. acknowledges support from grant AYA2013-48226-C3-1-P from the Spanish Ministry of Economy and Competitiveness (MINECO)
NRN is supported by the PRIN-INAF 2014 "Fornax Cluster Imaging and Spectroscopic Deep Survey" (PI. N.R. Napolitano). RS acknowledges support from Brain Korea 21 Plus Program(21A20131500002) and the Doyak Grant(2014003730).
Any opinion, finding and conclusion or recommendation expressed in this material is that of the author(s) and the NRF does not accept any liability in this regard.

\bibliography{biblio.firstmky.bib}

\end{document}